\documentclass[cup6b]{cupbook}
\usepackage{psfig}

\newcommand\sgra {SGR~1806$-$20}
\newcommand\sgrb {SGR~0526$-$66}
\newcommand\sgrc {SGR~1900$+$14}
\newcommand\sgrd {SGR~1627$-$41}
\newcommand\sgre {SGR~1801$-$23}                
\newcommand\sgrf {SGR~1808$-$20}                

\newcommand\axpa {1E~2259$+$586}
\newcommand\axpb {1E~1048.1$-$5937}
\newcommand\axpc {4U~0142$+$61}
\newcommand\axpd {1RXS~J170849.0$-$400910}
\newcommand\axpe {1E~1841$-$045}
\newcommand\axpf {XTE~J1810$-$197}
\newcommand\axpg {AX~J1845$-$0258}              
\newcommand\axph {CXOU~J0110043.1$-$721134}     

\def\lesssim{\mathrel{\hbox{\rlap{\hbox{\lower4pt\hbox{$\sim$}}}\hbox{$<$}}}}
\def\gtrsim{\mathrel{\hbox{\rlap{\hbox{\lower4pt\hbox{$\sim$}}}\hbox{$>$}}}}

\begin{document}

\setcounter{chapter}{13}

\pagenumbering{roman}

\pagenumbering{arabic}

\author[P.M. Woods and C. Thompson]{
P.M. Woods\\Universities Space Research Association\\National Space Science and
Technology Center\\Huntsville, AL, USA \and
C. Thompson\\Canadian Institute for Theoretical Astrophysics\\60 St.\
George Street\\ Toronto, ON, Canada}

\chapter[Soft Gamma Repeaters and Anomalous X-ray Pulsars\\]{Soft Gamma 
Repeaters and Anomalous X-ray Pulsars: Magnetar Candidates}

\vspace{2cm}

\section{Introduction}

Baade \& Zwicky (1934) were the first to envision the formation of neutron
stars as the end product of a supernova explosion.  Their forward thinking was
not vindicated for another three decades, with the discovery of the
first radio pulsars by Bell and Hewish (Hewish et al.\ 1968). 
What Baade and Zwicky could not have anticipated, however, was the menagerie of
astrophysical objects that are now associated with neutron stars.
Today, we observe them as
magnetically braking pulsars, accreting pulsars in binary systems, isolated
cooling blackbodies, sources of astrophysical jets, and emitters of
high-luminosity bursts of X-rays.  Here, we focus on two of the
most extraordinary evolutionary paths of a neutron star, namely Soft Gamma
Repeaters (SGRs) and Anomalous X-ray Pulsars (AXPs).

Soft Gamma Repeaters were discovered as high-energy transient burst sources;
some were later found also to be persistent X-ray pulsars, with periods
of several seconds, that are spinning down rapidly.  Anomalous X-ray Pulsars
are identified through their persistent pulsations and rapid spin down;
some have also been found to emit SGR-like bursts.  In spite of the
differing methods of discovery, this convergence in the observed
properties of the SGRs and AXPs has made it clear that they are,
fundamentally, the same type of object.  What distinguishes them
from other neutron stars is the likely source of energy for their
radiative emissions, magnetism.  The cumulative behavior of SGRs and AXPs 
is now best described by the magnetar model, in which the decay of 
an ultra-strong magnetic field ($B > 10^{15}$ G) powers the high-luminosity 
bursts and also a substantial fraction of the persistent X-ray emission.

For many years, the apparent absence of radio pulsars with 
magnetic fields much exceeding $10^{13}$ G, and the apparent lack of a
good motivation for the existence of much stronger fields in neutron 
stars, inhibited serious consideration of their astrophysical
consequences.  It was noted early on that fields as strong as 
$10^{14}-10^{15}$ G could be present in neutron stars as the result 
of flux conservation from the progenitor star (Woltjer 1964).  
Ultra-strong magnetic fields were introduced by hand in simulations 
of rotating supernova collapse, as a catalyst for energetic outflows 
(LeBlanc \& Wilson 1970; Symbalisty 1984).  A related possibility is
that ordinary radio pulsars could contain intense toroidal magnetic fields as 
a residue of strong differential rotation in the nascent neutron star (e.g.
Ardelyan et al.\ 1987).  Later it was realized that appropriate conditions
for true dynamo action could exist in proto-neutron stars (Thompson
\& Duncan 1993), leading to the formation
of a class of ultra-magnetic neutron stars with dissipative properties 
distinct from those of radio pulsars (Duncan \& Thompson 1992).
In recent years, pulsar searches have largely 
closed the observational gap between the dipole fields of radio pulsars and 
magnetar candidates (Manchester 2004). 

We first review the history of this relatively new subfield
of high energy astrophysics.  Then we summarize in more detail
the burst emission, the persistent X-ray emission of the SGRs and AXPs,
their torque behavior, the counterparts observed in other wavebands, 
and their associations with supernova remnants.  Finally, we discuss 
the magnetar model.

\subsection{Soft Gamma Repeaters: A brief history}

On 1979 January 7, a burst of soft gamma-rays lasting a quarter of a second was
detected from \sgra\ by instruments aboard the {\it Venera} spacecraft -- the
first observation of a Soft Gamma Repeater.  This burst, along with a handful
like it recorded over the next few years, were originally classified as a
subtype of classical Gamma-Ray Burst (GRB), one with a short duration and a
soft spectrum  (Mazets \& Golenetskii 1981).  The locations of three repeaters 
were obtained from this early data set.  It was not until after an intense 
reactivation of \sgra\ in 1983, however, that the independent nature of these
sources was fully appreciated (Hurley 1986; Laros et al.\ 1987).  Their
propensity to emit multiple bursts (no GRB has yet been shown to repeat); the
deficit of high energy gamma-ray emission; and their similarity to each other
merited designation as a new class of astrophysical transient.

The first detection of a SGR burst was soon followed,
on 1979 March 5, by the most energetic SGR flare yet recorded
(Mazets et al.\ 1979).
This extraordinary event began with an extremely bright spike
peaking at $\sim$10$^{45}$\,ergs s$^{-1}$ (Golenetskii et al.\ 1984),
followed by a 3 minute train of coherent 8\,s pulsations whose flux 
decayed in a quasi-exponential manner (Feroci et al.\ 2001).  
The burst was well localized at the edge of the supernova remnant (SNR)
N49 in the Large Magellanic Cloud, and its source is now identified
as \sgrb\ (Cline et al.\ 1982).  The high
luminosity, strong pulsations, and apparent association with a SNR strongly
suggested that the source was a young, magnetized neutron star
with a spin period of 8\,s. 

Following the announcement of the SGRs, a variety of models were proposed. 
These included accretion onto magnetized neutron stars  (e.g.\ Livio \& Taam
1987; Katz, Toole \& Unruh 1994), cometary accretion  onto quark stars (Alcock,
Farhi \& Olinto 1986), as well as thermonuclear  energy release on a magnetized
neutron star (Woosley \& Wallace 1982).   Damping of the vibration of a neutron
star had been suggested as a mechanism for the March 5 flare (Ramaty et al.\
1980), but the coupling of the crust to the magnetosphere is much too weak to
explain the observed luminosity if $B \sim 10^{12}$ G (Blaes et al.\ 1989).
Indeed, the main shortcoming of all these models (e.g.\ Norris et al.\ 1991)
was the lack of an adequate explanation for both the giant flare and the more
common recurrent bursts, which last only $\sim$0.1 s and have much lower peak
luminosities ($<$10$^{41}$ ergs s$^{-1}$).  

Efforts to understand the nature of the SGRs were constrained by 
the lack of information about persistent counterparts.   This changed 
with the discovery of persistent X-ray emission from
all three known SGRs (Murakami et al.\ 1994; Rothschild, Kulkarni
\& Lingenfelter 1994; Vasisht et al.\ 1994).  Around the same time,
the magnetar model was put forth to explain the high-luminosity
bursts of the SGRs (Duncan \& Thompson 1992; Paczy\'nski 1992; 
Thompson \& Duncan 1995) and the persistent X-ray emission
of both the SGRs and the AXPs (Thompson \& Duncan 1996).  Thompson \& Duncan
(1996) predicted slow pulsations and rapid spin down from the quiescent X-ray
counterparts of the SGRs.  A major breakthrough in determining the nature of
SGRs was made shortly thereafter (Kouveliotou et al.\ 1998a), with the 
discovery  of 7.5 s pulsations and rapid spin down in the X-ray counterpart 
to \sgra.  Kouveliotou et al.\ (1998a) interpreted this measurement
in terms of the magnetic braking of an isolated neutron star with a
$\simeq 10^{15}$ G dipole magnetic field.  

Our understanding of SGRs has continued to blossom in recent years in good part
due to extensive monitoring campaigns and improved instrumentation. These
observations have revealed  correlated changes in SGR persistent emission
properties during periods of burst activity, dramatic variations in spin down
torque, and a much larger collection of bursts of all types.  For example, a
near carbon-copy of the first giant flare was recorded on 1998 August 27 from
\sgrc\ (Hurley et al.\ 1999a).  In spite of detector advancements, there has
been only one additional confirmed SGR discovered after the first three in
1979:  \sgrd\ emitted more than 100  bursts in 1998 (Kouveliotou et al.\ 1998b;
Woods et al.\ 1999a).  Two bursts were recorded in 1997  from a fifth candidate
source, \sgre\ (Cline et al.\ 2000).  Another  candidate, \sgrf, was detected
once and localized (Lamb et al.\ 2003a) to a position very near, but formally
inconsistent with, the direction of \sgra.  It should be cautioned that this 
burst was recorded during a burst active phase of \sgra.   Note that we  have
not included in this tally sources first identified as AXPs, and  later found
to burst like SGRs.

\subsection{Anomalous X-ray Pulsars: A brief history}

The first detection of an Anomalous X-ray Pulsar was made by Fahlman \& Gregory
(1981), who discovered pulsations from the X-ray source \axpa\ at the center of
the SNR CTB~109.  This object was first interpreted as a peculiar X-ray
binary: its energy spectrum 
was much softer than is typical of accreting  pulsars, and no optical 
counterpart was detected.  Later the source was found to be spinning down
in a secular manner (Koyama, Hoshi \& Nagase 1987).  Its X-ray 
luminosity was much too high to be powered by the loss of rotational 
energy from the putative neutron star. 

Several other similar sources were discovered in the ensuing fifteen years.
The objects \axpa, \axpb, and \axpc\ were grouped together by Hellier (1994) 
and Mereghetti \& Stella (1995) as possible low-mass X-ray binaries, along 
with the known short-period binary 4U~1626$-$67.  (The source 
RX~J1838.4$-$0301 was also included initially, but was later shown not to be
an X-ray pulsar.)  The salient properties of this class were 
a narrow range  of spin periods (5$-$9 s), 
fairly constant X-ray luminosities ($\sim$10$^{35}$$-$$10^{36}$ 
ergs s$^{-1}$), no evidence for orbital Doppler shifts and --
with the exception of 4U~1626$-$67 -- relatively soft X-ray spectra and
steady spin down.  However, 4U~1626$-$67 is also distinguished from the other
sources by the detection of optical pulsations of the brightness expected
from a compact, accreting binary.  In light of these differences, its
membership as an AXP has been revoked.  Three new AXPs 
have been discovered since 1996 (\axpd, \axpe, 
and \axpf), along with two candidate sources (\axpg\ and 
CXOU J0110043.1$-$721134).

The AXPs nonetheless appear to be too young to be low-mass binaries:  
some are associated with SNR, and they have a small scale height above 
the Galactic plane (van Paradijs, Taam \& van den Heuvel 1995).
As noted by Thompson \& Duncan (1993, 1996), their `anomalous'
property is the mechanism powering their X-ray emission.  These
authors identified \axpa, and later the AXP population as a whole, with
isolated magnetars powered by the decay  of a $\sim 10^{15}$ G magnetic field.
The principal competing model postulated that
the AXPs are neutron stars surrounded by fossil disks that were
acquired during supernova collapse or during a common-envelope
interaction (Corbet et al.\ 1995; van Paradijs et al.\ 1995;
Chatterjee, Hernquist, \& Narayan 2000; Chatterjee \& Hernquist 2000).
Finally, it was also noted that the loss of
rotational energy from an isolated, magnetic, high-mass white dwarf
is much larger than from a neutron star with the same spin parameters,
and could supply the observed X-ray output (Paczy\'nski 1990).
However, the apparent youth of the object, and its residence in a SNR,
remained puzzling in that interpretation.  

The detection of optical and near infrared counterparts to the AXPs, 
beginning with \axpc\ (Hulleman, van Kerkwijk \& Kulkarni 2000), has 
provided a useful discriminant between the fossil disk and magnetar models.
Dim counterparts have now been detected for four AXPs, with
an optical/IR luminosity typically one thousandth of that emitted in
2$-$10 keV X-rays.
This constrains any remnant accreting disk to be very compact
(e.g.\ Perna, Hernquist \& Narayan 2000).  The optical emission of \axpc\ 
has been found to pulse at the same period as in the X-ray band, with a pulsed
fraction that is equal or higher (Kern \& Martin 2002).  The large
pulsed fraction appears problematic in any accretion model, where
the optical emission arises from re-processing of the X-rays by a disk.
There are no reliable {\it a priori} predictions of optical/infrared 
emission from magnetars.

The detection of X-ray bursts similar to SGR bursts from at least one, and
possibly two, AXPs has confirmed a key prediction of the magnetar model.   Two
weak bursts were observed from the direction of \axpb\ (Gavriil, Kaspi \& Woods
2002); and more than 80 SGR-like bursts were detected from \axpa\ during a
single, brief ($\sim$11 ks) observation of the source (Kaspi et al.\ 2003). 
Overall, at least 10 percent of the X-ray output of \axpa\ appears to be
powered by transient releases of energy, and a much larger fraction in some
other AXPs.   Although these observations have not yet provided unambiguous
proof that the AXPs have  ultra-strong magnetic fields, they have confirmed the
conjecture that  the AXPs and SGRs belong to the same class of neutron stars.  
In this review, we refer to these sources collectively as magnetar candidates.

\section{Burst Observations}

The defining behavior of SGRs is their repetitive emission of bright bursts of
low-energy (soft) gamma-rays.  The most common SGR bursts have short  durations
($\sim 0.1$ s), thermal spectra, and peak luminosities reaching up to $10^{41}$
ergs s$^{-1}$ --- well above the standard Eddington limit of $\sim 2 \times
10^{38}$ ergs s$^{-1}$ for a 1.4 $M_{\odot}$ neutron star.  In this section, we
describe these short bursts.  We include the very similar bursts detected 
from two AXPs, which turn out to be remarkably similar to the SGRs
bursts in terms of their durations, spectra, and energy distribution
(Gavriil, Kaspi, \& Woods 2004).
We then review the more extraordinary bursts emitted by the SGRs, including the
two giant flares and the intermediate bursts.  Unless otherwise stated,
the quoted burst luminosities and energies cover photon energies above 
20 keV and assume isotropic emission.


\begin{figure}[!htb]

\centerline{
\psfig{file=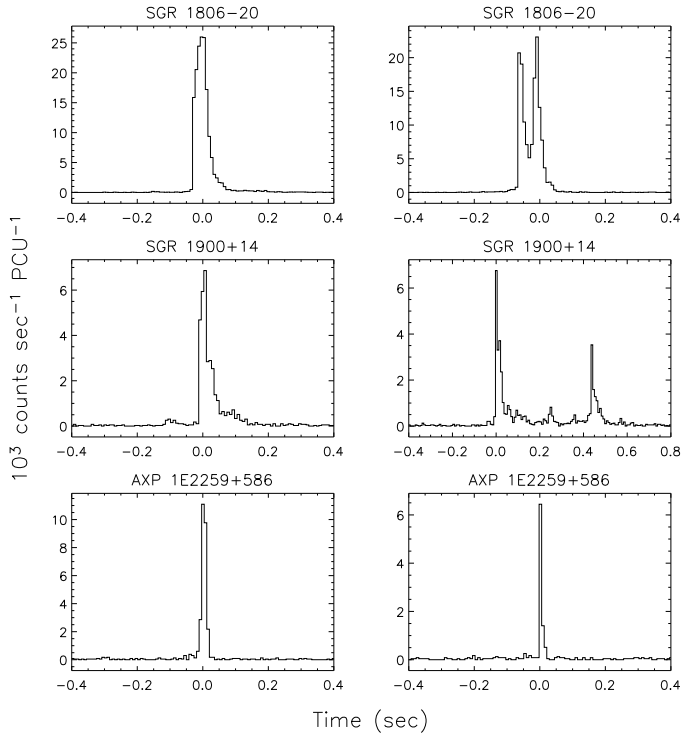,height=4.0in}}

\caption{A selection of common burst morphologies recorded from \sgra, \sgrc\
and \axpa, as observed with the {\it RXTE} PCA.  All light curves display
counts in the energy range 2$-$20 keV, with a time resolution of 7.8 ms.
See text for further details.\label{fig:burst_selection}}

\end{figure}

\subsection{Short Duration SGR Bursts: Temporal Properties and 
Energy Distribution}

The properties of the most common SGR bursts do not appear to vary greatly
between different periods of activity, or indeed between different sources
(e.g.\ Aptekar et al.\ 2001; {G\"o\u{g}\"u\c{s}} et al.\ 2001). A burst
typically has a faster rise than decay, and lasts $\sim100$ ms.  Four examples
from \sgra, \sgrc, and \axpa\ are shown in Figure~\ref{fig:burst_selection}.  
A number of bursts are multi-peaked, like the two shown from \sgra\  and
\sgrc.  Complicated bursts like these can usually be decomposed into burst
``units.''  {G\"o\u{g}\"u\c{s}} et al.\ (2001) showed that the intervals
between sub-peaks have a broad distribution, suggesting that these multi-peaked
bursts are a superposition of two (or more) single-peaked burst units close in
time.


\begin{figure}[!htb]

\centerline{
\psfig{file=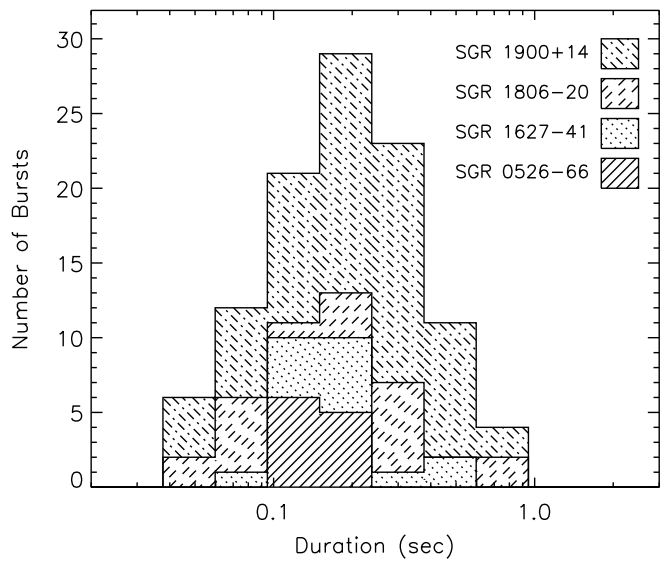,height=3.0in}}

\caption{Duration distribution for 106 bursts from the four known SGRs as
observed by the Konus detectors (15$-$100 keV) between 1978 and 2000 (Aptekar
et al.\ 2001).\label{fig:dur_dist}}

\end{figure}

The morphological uniformity of (the majority of) SGR bursts was noted
early on (e.g.\ Atteia et al.\ 1987; Kouveliotou et al.\ 1987).  The burst
durations have a narrow distribution:  they show a mild positive correlation
with burst fluence (e.g., {G\"o\u{g}\"u\c{s}} et al.\ 2001), but do not vary
significantly with photon energy.
A sample of 164 bursts recorded from the four known SGRs by the
Konus series of gamma-ray detectors is tabulated by Aptekar et al.\ (2001).
Durations could be measured for 106 ordinary bursts
(Figure~\ref{fig:dur_dist}), with a mean of 224 ms.   More recently, the higher
flux sensitivity of the {\it Rossi X-ray Timing Explorer} PCA has provided
larger burst samples for individual sources, at lower  photon energies (2$-$20
keV vs. $> 25$ keV).  In particular, $T_{90}$ burst durations\footnote{The time
to accumulate 90 percent of the burst fluence; see Koshut et al.\ (1996).} were
measured for 190, 455, and 80 bursts from the magnetar candidates \sgra, \sgrc\
({G\"o\u{g}\"u\c{s}} et al.\ 2001), and \axpa\ (Gavriil et al. 2004),
respectively.  The mean durations of these samples were 162, 94, and 99 ms.
That these values are somewhat lower than in the Konus sample may be due, in
part, to the  higher mean fluence of the Konus bursts.


\begin{figure}[!htb]

\centerline{
\psfig{file=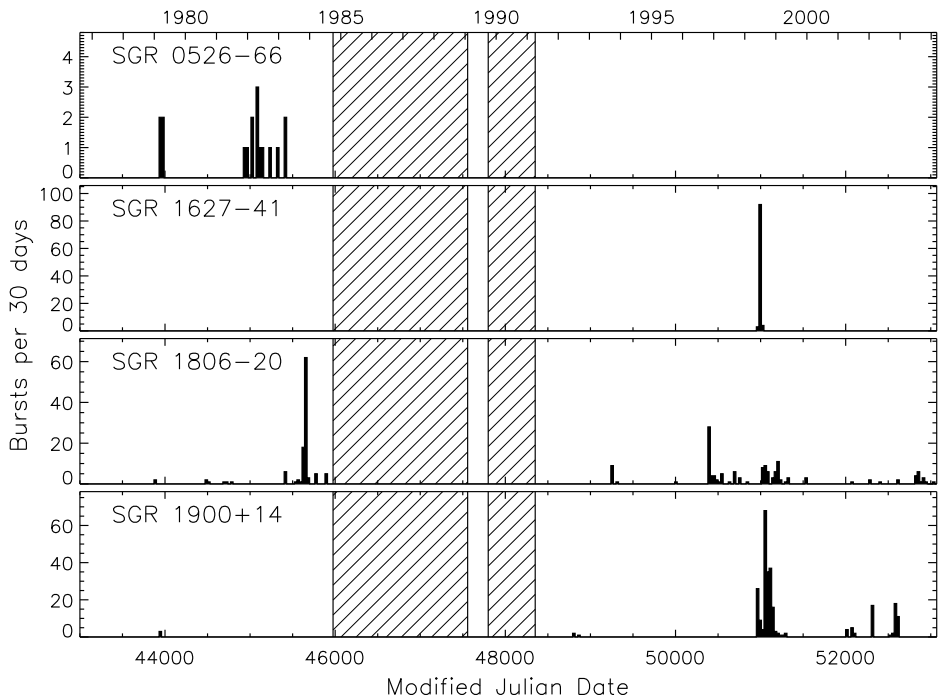,width=5.0in,angle=90}}
\vspace{-0.1in}

\caption{Burst activity history of the four confirmed SGRs.  The bursts
identified here were detected with a suite of large field-of-view
detectors having different sensitivities.  The shaded regions indicate epochs
where there were no active detectors sensitive to SGR bursts.  IPN data
courtesy of Kevin Hurley.\label{fig:sgr_bhist}}

\end{figure}

The burst activity in SGRs tends to be concentrated in time.  These episodes of
enhanced burst activity are referred to as outbursts.  They occur at irregular
intervals with variable duration and intensity (Figure~\ref{fig:sgr_bhist}).
Within each outburst, the
recurrence patterns of individual bursts are just as irregular as those
of the outbursts themselves, and differ dramatically from what is observed 
in X-ray bursts (of either Type I or II) in accreting neutron stars (\S3).  
There
is no correlation between the energy of a given burst and the time to the 
next burst in either the SGRs (Laros et al.\ 1987; {G\"o\u{g}\"u\c{s}} et 
al.\ 1999), or in \axpa\ (Gavriil et al.\ 2004).  (Such a correlation
is present in the Type II bursts of the Rapid Burster; see, for example,
Lewin, van Paradijs, \& Taam 1993).  

The distribution of waiting times between bursts follows a log-normal function
with a mean that depends on the sensitivity of the detector and the strength of
the outburst.  For example, the waiting times  spanned some 7 orders of
magnitude during the 1983 activation of \sgra\ with a (logarithmic) mean of
$\sim$10$^4$ s (Laros et al.\ 1987; Hurley et al.\ 1994).  Cheng et al.\ (1986)
pointed out that the waiting times between earthquakes show a similar
distribution. The waiting times between bursts from \sgrc, \axpa,  and a more
recent outburst of \sgra\ are all consistent with a log-normal distribution,
although given the lower flux threshold the mean waiting time is only  $\sim
10^2$ s in these three samples.

The energies radiated during the common ($\sim 0.1$-s) SGR bursts follow  a
power-law distribution, $dN/dE \propto E^{-5/3}$.  Cheng et al.\ (1996) first
uncovered this distribution in \sgra, and pointed out the similarity with the
Gutenburg-Richter law for earthquakes.  Similar distributions are measured in a
variety of other physical systems, including Solar flares and avalanches. 
Subsequently, it has been shown that the other three SGRs and the AXP \axpa\
all possess very similar burst energy distributions.  The power-law index is
about $-5/3$ in \sgrc, \sgrd, and \axpa, but is not well constrained in \sgrb. 
A possible break from a $-5/3$ index  to a somewhat flatter value ($-1.4$) at
low burst energies was measured in a larger sample of bursts from \sgra\
({G\"o\u{g}\"u\c{s}} et al.\ 1999).

\begin{figure}[!htb]
\centerline{
\psfig{file=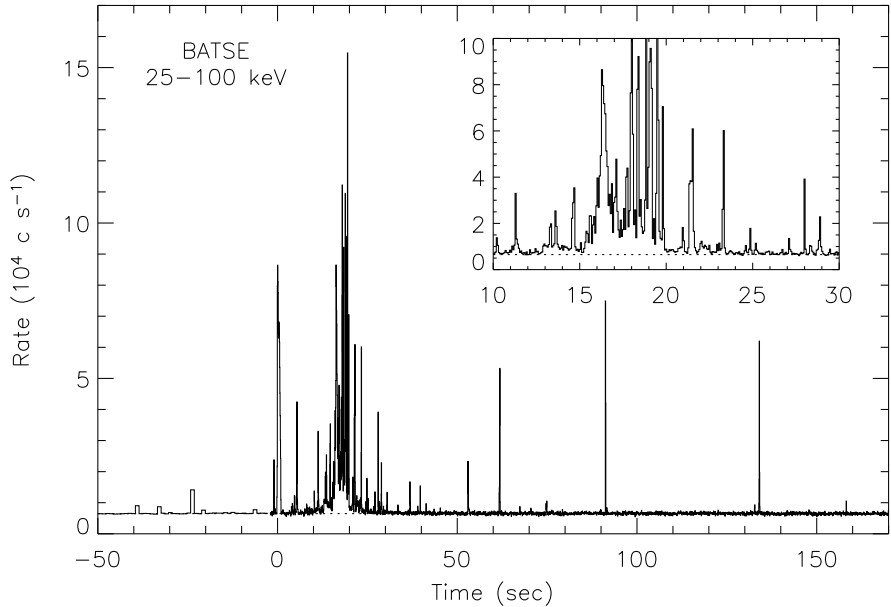,height=3.2in}}
\vspace{-0.05in}

\caption{The time history of the multi-episodic burst from SGR~1900$+$14
recorded on 1998 September 1 with BATSE (25$-$100 keV).  The inset shows a
close-up of the most intense part of the light curve.  The background level
is indicated by the dotted line.  Note the envelope of emission lasting
$\sim$5$-$7 s during the most intense phase.
\label{fig:burst_sep01}}

\vspace{11pt}
\end{figure}

Series of many short bursts, with extremely small waiting times
(multi-episodic bursts) have been observed on rare occasions
(Hurley et al.\ 1999b).   They involve several tens of bright SGR bursts
which are packed into an interval of a few minutes.
Intense burst episodes like these are more commonly seen at lower
peak flux; but three instances involving high luminosity SGR bursts have been
recorded from \sgrc. The BATSE light curve of the 1 September 1998
multi-episodic burst is  shown in Figure~\ref{fig:burst_sep01}.  Note the
continuous envelope of emission underlying the most intense portion of the
burst episode.



\subsection{Spectral Properties}

Bursts from SGRs were discovered using all-sky detectors with little
sensitivity below $\sim 30$ keV.  Above this photon energy, SGR burst spectra
are well modeled by optically thin thermal bremsstrahlung (OTTB).  The
temperatures so obtained fall within the narrow range $kT = 20-40$ keV,
indicative of the spectral uniformity of SGR bursts.  The spectra of SGR bursts
vary weakly with intensity -- not only from burst to burst within a given
source, but also between sources.  This effect was first demonstrated by
Fenimore, Laros, \& Ulmer (1994) for \sgra, and later by Aptekar et al.\ 
(2001) in the Konus sample of bursts from four SGR sources.


\begin{figure}[!htb]

\centerline{
\psfig{file=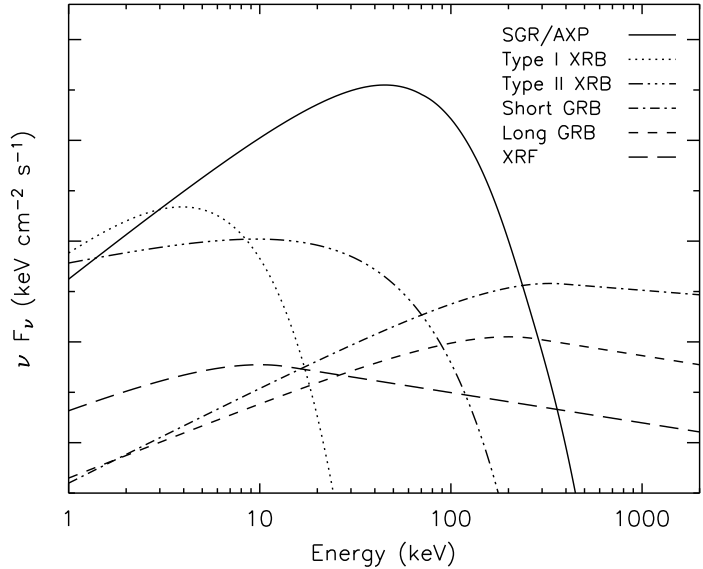,height=3.2in}}

\caption{Characteristic energy spectra of high-energy burst phenomena.  The
SGR/AXP (magnetar candidate) burst spectrum shown is a composite of the two
blackbody model that fits burst spectra well below $\sim$50 keV convolved  with
the OTTB model that better represents the burst spectrum at higher  energies. 
Note that there exists a continuum of $E_{\rm peak}$ values between the softer
XRFs and the harder/brighter long GRBs. \label{fig:burst_spec}}

\end{figure}

A typical SGR burst spectrum (the solid line in Figure~\ref{fig:burst_spec}) is
compared with sample spectra from other  extra-Solar high-energy burst
phenomena, specifically, short Gamma-Ray Bursts, long Gamma-Ray Bursts, X-ray
Flashes (XRFs), thermonuclear burning or Type I X-ray bursts (XRBs), and
spasmodic accretion or Type II X-ray bursts (example shown is a GRO~J1744$-$28
burst spectrum [e.g.\ Giles et al.\ 1996] -- Rapid Burster energy spectra are
significantly softer [Lewin et al.\ 1993]).  Although the soft end of the
spectral distribution of  GRBs and X-ray Flashes (XRFs) overlaps the SGRs, the
durations of the SGR bursts  are usually shorter by two orders of magnitude
than those of the detected X-ray Flashes (whose durations are $\gtrsim 10$ sec)
and otherwise show strong morphological differences with GRBs.  Distinguishing
the two types of bursts is therefore straightforward in practice.  For reviews
on XRFs and GRBs, see (\S6) and (\S15). One caveat here is that the initial
$\sim 0.3$-s spikes of the giant flares show greater spectral  similarities
with GRBs, and so a handful of extra-galactic SGR flares may be hidden in the
BATSE catalog of short-duration GRBs (Duncan 2001).

A shortcoming of the OTTB model is that it over-predicts the
flux of photons with energies below $\sim 15$ keV (Fenimore et al.\ 1994).  
It is doubtful that this spectral rollover is due to a thick column
of absorbing material, since the requisite $N_{\rm H}$ is an 
order of magnitude greater than what is deduced from the persistent
X-ray emission.  Recently,
a $7-150$ keV {\it HETE-2} spectrum of a high-fluence burst from  \sgrc\ was
successfully fit by the sum of 4.1 keV and 10.4 keV blackbodies (Olive et al.\
2003).  A similar result was obtained with $1.5-100$ keV BeppoSAX spectra of 10
bursts also from \sgrc\ (Feroci et al.\ 2004).  The temperatures of these lower
peak flux bursts are consistent with the {\it HETE-2} burst spectrum --
so that the flux ratio of the two blackbody components is approximately
constant.  Furthermore, the absorbing column measured during the bursts 
is consistent the value obtained in quiescence.

The improved sensitivity of {\it RXTE} allowed {G\"o\u{g}\"u\c{s}} et al.\
(2001) to show that the less energetic bursts from \sgra\ and \sgrc\ are also
slightly harder spectrally.  The bursts detected from the AXP \axpa\
have similar spectra to those of the SGRs, although in the AXP it is the 
brighter bursts which tend to be harder (Gavriil et al.\ 2004).

\subsection{Giant Flares}

Giant flares are the most extreme examples of SGR bursts.  Their output
of high energy photons is exceeded only by blazars and cosmological 
gamma-ray bursts, and their luminosity peaks above a million times
the Eddington luminosity of a neutron star.  The flares begin with a 
$\sim 1$ second spike of spectrally hard emission which decays rapidly 
into a softer, pulsating tail that persists for hundreds of seconds.  These
coherent pulsations are at the spin period of the underlying neutron star.
The giant flares are rare:  only two have been detected from the 
four known SGRs over 20 years of observation, so the corresponding
rate is approximately once per 50$-$100 yr (per source).  In contrast
with GRBs and blazars, there is no evidence for strong beaming in the
SGR bursts.

The first giant flare was recorded on 1979 March 5 from \sgrb\ (Mazets et al.\
1979) and, indeed, was only the second SGR burst observed.  The source is well
localized in the LMC (\S14.6), and so the isotropic energy of the flare was
$5\times 10^{44}$ ergs -- some ten thousand times larger than a typical
thermonuclear flash.  The initial peak of this flare lasted $\sim 0.2$ s and
had significant structure on time scales shorter than $\sim 2$ ms.   It was
spectrally harder ($kT \sim 250-500$ keV) than the common SGR bursts, and
reached a peak luminosity of $4\times 10^{44}$  ergs s$^{-1}$ (Mazets et al.\
1979; Fenimore et al.\ 1981).  Thereafter, the flux decayed in a
quasi-exponential manner over the next $\sim$2-3 minutes.  A reanalysis of the
{\it ISEE-3} data using a model of a magnetically confined, cooling fireball
(see \S14.7.2), shows that the data are also consistent with a well-defined
termination of the X-ray flux at $\sim 160$ s (Feroci et al.\ 2001).  The
pulsations during this phase of the burst have a period of $8.00 \pm 0.05$ s 
(Terrell et al.\ 1980).  The pulse profile shows two clear peaks per cycle and
a change in morphology during the first few cycles.   The spectrum of the
decaying tail had an OTTB temperature of $\sim 30-38$  keV, consistent with the
spectra of the recurrent burst emissions  from this SGR.  



\begin{figure}[!htb]

\centerline{
\psfig{file=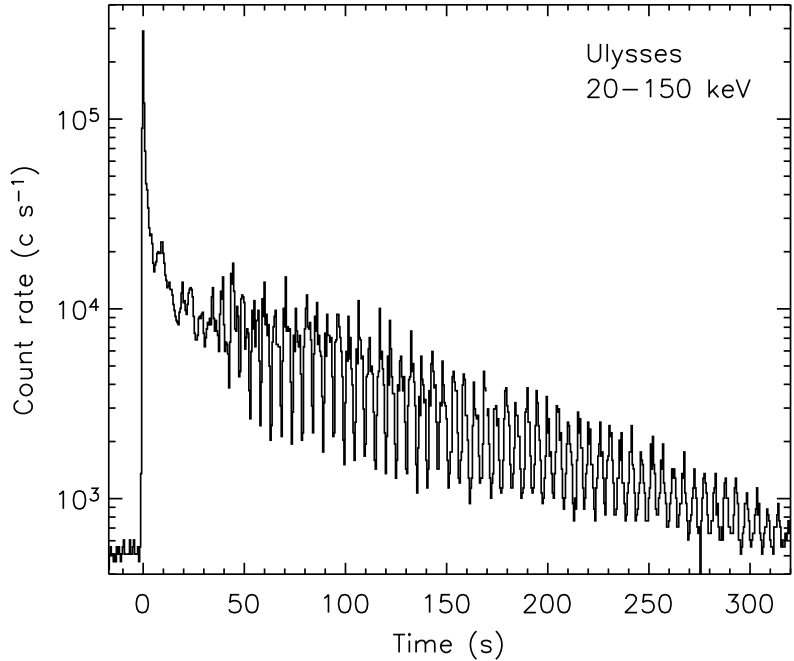,height=3.5in}}

\caption{The giant flare from SGR~1900$+$14 as observed with the gamma-ray
detector aboard Ulysses (20$-$150 keV).  Note the strong 5.16 s pulsations
clearly visible during the decay.  Figure after Hurley et al.\ (1999a).
\label{fig:aug27}}

\end{figure}

The second giant flare was not recorded until almost 20 years later, on 1998
August 27 from \sgrc\ (Hurley et al.\ 1999a; Feroci et al.\ 1999; Mazets et
al.\ 1999a; Feroci et al.\ 2001).  This event (Figure~\ref{fig:aug27}) was, in
many respects, a carbon copy of the March 5 flare.  It began with a bright
spike lasting  $\sim 0.35$ s, and the X-ray spectrum contained a very hard
power-law  component $dN/dE \propto E^{-1.5}$ in the initial stages.  Only a
lower  bound of $3 \times 10^{44}$ ergs s$^{-1}$ was obtained for its peak
luminosity, because the flare saturated every detector that observed it.  In
fact, it was the brightest extra-Solar gamma-ray transient yet recorded.  The
X-ray flux incident on the night side of the Earth was high enough to force
the ionosphere to its day-time level (Inan et al.\ 1999).  Its total energy
exceeded $10^{44}$ ergs.

In contrast with the previous flare, the decline in the flux from the  August
27 flare was followed to a well-defined termination some 400 s after the
initial spike (Feroci et al.\ 2001).  The spectrum, after the first 50 s,
equilibrated to a (OTTB) temperature of $\sim 30$ keV, even while the
luminosity continued to decrease by more than an  order of magnitude.  During
this same phase, the light curve maintained large-amplitude pulsations with a
5.16 s period, precisely equal to the  periodicity that had been previously
detected in the persistent X-ray  emission of \sgrc\ (Hurley et al.\ 1999c). 
The pulse maintained a complex  four-peaked pattern that gradually simplified
into a smooth single  pulse during the final stages of the flare (Mazets et
al.\ 1999a).

\subsection{Intermediate Bursts}

Intermediate bursts are intermediate in duration, peak luminosity and  energy
between the common recurrent SGR bursts and the giant flares. They have
durations of seconds or longer, and peak luminosities exceeding $\sim 10^{41}$
ergs s$^{-1}$.  They tend to have abrupt onsets and, if the duration is less
than the rotation period of several seconds, also abrupt end points. The flux
generally varies smoothly in between. The short, recurrent bursts (e.g.\
{G\"o\u{g}\"u\c{s}} et al.\ 2001)  are usually more irregular, which suggests
that the emitting particles  cool more rapidly.  The intermediate bursts are
most commonly observed in the days and months following the giant flares,
which  suggests that they represent some residual energy release by essentially
the same mechanism.  In a nutshell, these bursts appear to be ``aftershocks''
of the giant flares.



\begin{figure}[!h]

\centerline{
\psfig{file=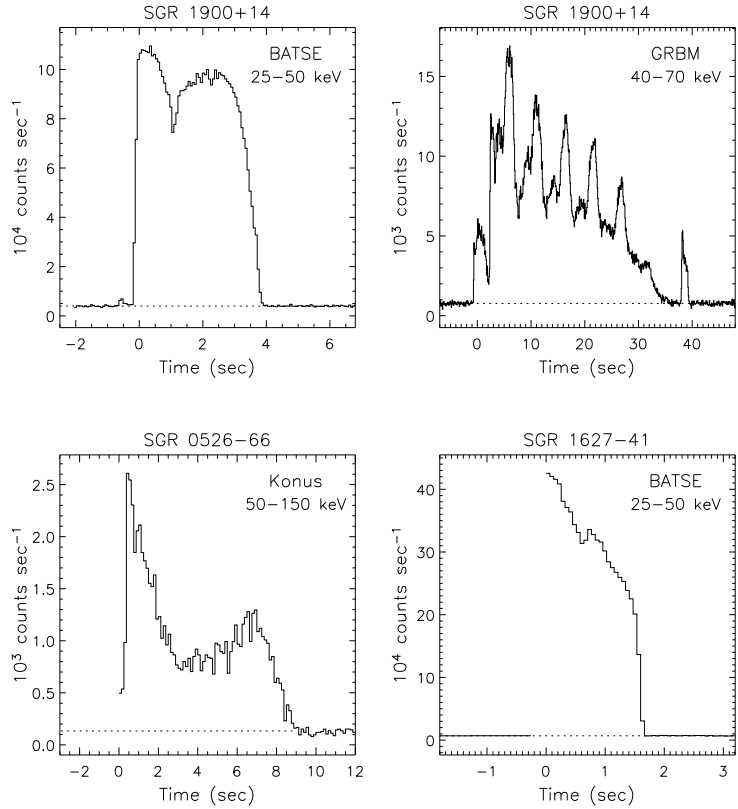,height=4.3in}}
\vspace{-0.15in}

\caption{Time histories of four intermediate bursts recored from three of the
four SGRs.  Clockwise from upper left: SGR~1900$+$14 burst recorded with BATSE
on 1998 October 28, SGR~1900$+$14 burst recorded with GRBM on 2001 April 18,
SGR~1627$-$41 burst recorded with BATSE on 1998 June 18, and SGR~0526$-$66
burst recorded with Konus on 1982 February 27.  Energy ranges are shown in each
figure panel.  The rise of the SGR~1627$-$41 burst is unresolved due to a gap
in the BATSE data.  GRBM data courtesy of M.~Feroci and F.~Frontera.  Konus
data courtesy of S.~Golenetskii. \label{fig:bigbursts}}

\end{figure}

Time histories of
four examples of intermediate bursts are shown in  Figure~\ref{fig:bigbursts}. 
Their isotropic energies range from 10$^{41}$$-$10$^{43}$ ergs.  
Up until 1998, there were
few intermediate bursts recorded from SGRs, most from \sgrb\ (Golenetskii
et al.\ 1984).  Since 1998, several more bursts have been detected from other
SGRs which begin to fill in the apparent gap in energy and duration. (The
largest was an event recorded on 2001 April 18 from \sgrc; Guidorzi et al.\
2004.)  This suggests that there may be a continuum of burst sizes covering the
smallest recurrent bursts all the way up to the giant flares.  



\begin{figure}[!htb]
\centerline{
\psfig{file=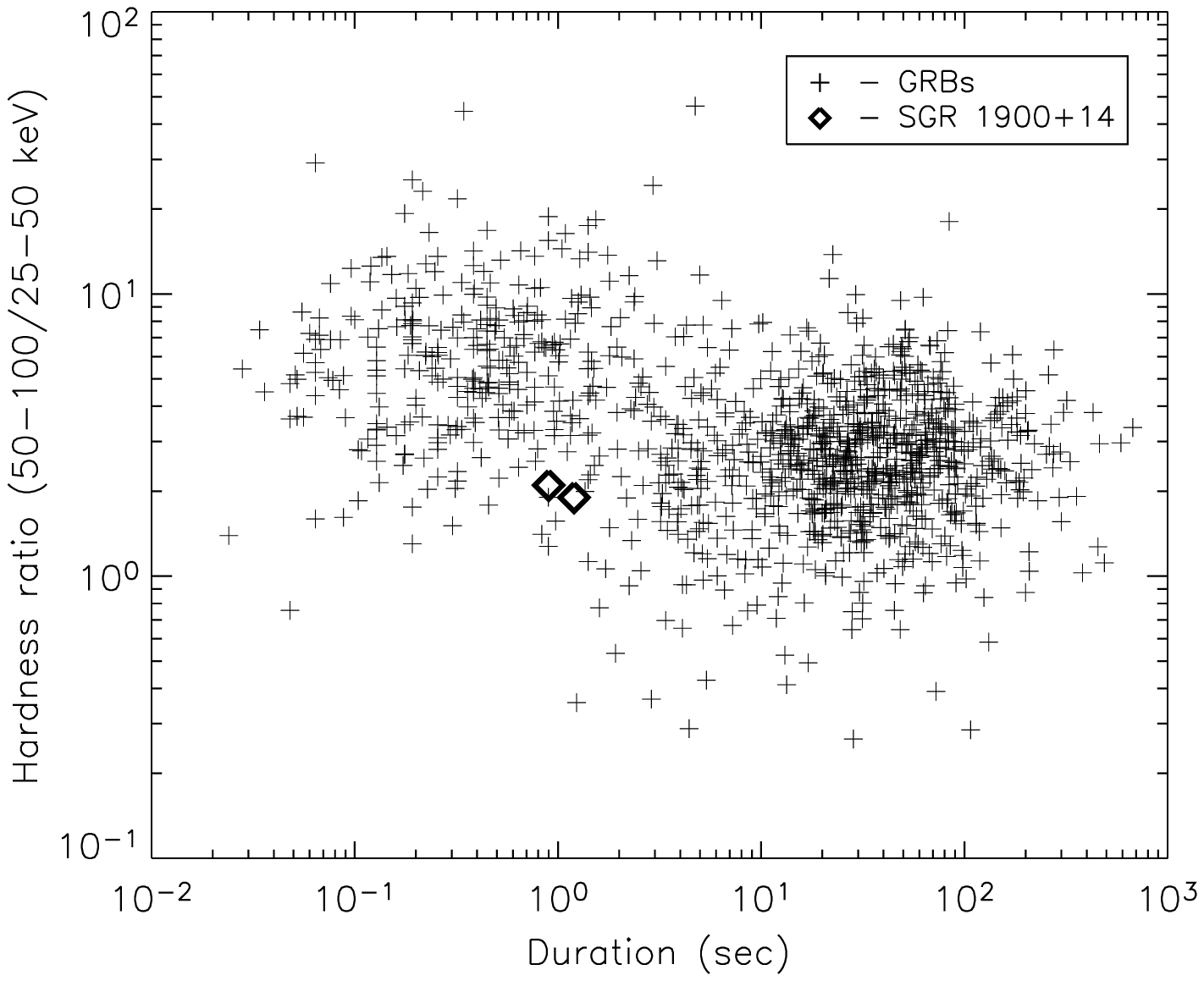,height=3.5in}}
\vspace{-0.05in}

\caption{Hardness ratio versus $T_{90}$ duration for all GRBs (plus signs) in
the BATSE 4B catalog (Paciesas et al.\ 2000) and the two spectrally hard \sgrc\
bursts (diamonds) detected with BATSE (Woods et al.\ 1999b).
\label{fig:hard_bursts}}

\vspace{11pt}
\end{figure}

The spectra of most intermediate bursts are consistent with the spectra of the
short, recurrent bursts and the pulsating tails of the giant flares.  The
spectra do not vary much, either from burst to burst or within individual
bursts.   A striking exception to this rule was a very intense ($L_{\rm peak}
\sim$10$^{43}$ ergs s$^{-1}$) and spectrally hard ($kT_{\rm peak}$ $\sim$120
keV) burst detected from \sgrd\ (Mazets et al.\ 1999b; Woods et al.\ 1999a). 
This burst lasted $\sim$0.5 s and was similar both spectrally and temporally to
the initial peaks of the giant flares -- but without the extended softer
pulsations. Two bursts recorded from  \sgrc\ during the 1998$-$1999 activation
were also spectrally much harder  than all other burst emission from this SGR
(Woods et al.\ 1999b) with  the exception of the initial spike of the August 27
flare.  These bursts  are, in fact, spectrally and temporally indistinguishable
from classical  GRBs (Figure~\ref{fig:hard_bursts}).  They were not
exceptionally bright and had durations  lasting $\sim$1 s with a fast rise and
exponential decay.  Their spectra were consistent with a power law (photon
index $\sim -2$) whose hardness was anti-correlated with X-ray flux.

\subsection{Possible Spectral Features}

Discrete features in burst spectra from magnetar candidates have been reported
from \sgrb, \sgrc, \sgra, and \axpb.  It should be emphasized that, as was the
case previously with classical gamma-ray bursts,  the  same spectral feature
has not yet been detected in the same burst by independent instruments.  

In \sgrb, Mazets et al.\ (1979) reported evidence for a broad peak in the
energy spectrum at $\sim$430 keV during the main peak of the giant flare of
March 5.  Using {\it RXTE} PCA data, Strohmayer \& Ibrahim (2000) discovered a
significant emission feature at $\sim$6.7 keV during a pre-cursor to the
intermediate burst of 1998 August 29 from \sgrc.  An additional feature
consistent with twice this energy is seen, but its significance is marginal.
Ibrahim, Swank \& Parke (2003) presented the analysis of 56 spectra accumulated
with the {\it RXTE} PCA taken from selected \sgra\ bursts intervals.  Of the 56
spectra, a handful showed a statistically significant ($>3\sigma$) absorption
feature near 5 keV and much less significant features at integer multiples of
this energy.  These authors have argued that these lines represent proton
cyclotron absorption features in a strong magnetic field. In addition,  two
bursts were recorded with the PCA from the direction of \axpb\ within two weeks
of each other late in 2001 (Gavriil et al.\ 2002).   In the first of these
bursts, a strong emission feature was seen at $\sim$7 keV with less significant
features at energies consistent with the first three harmonics.

\section{Persistent X-ray Emission}

Historically, one of the defining properties of AXPs was the relative
steadiness of their X-ray emission, over a fairly narrow range
10$^{35}-10^{36}$ ergs s$^{-1}$.  Over the last several years, however, it has
become clear that at least half and possibly most magnetar candidates are 
variable X-ray sources.  Some of the observed variability is clearly driven by
burst activity (see \S14.5), but at least a few sources have shown large
changes in luminosity ($\sim$10$-$100) with little or no detected burst
activity.  For example, \axpf\ was discovered  in 2003 at a luminosity of
$\sim$2 $\times$ 10$^{36}$ ergs s$^{-1}$  (Ibrahim et al.\ 2004), but archival
observations from the 1990's found  the source in a ``low state'' with a
luminosity two orders of magnitude smaller (Gotthelf et al.\ 2004).  One of the
AXP candidates, \axpg, was discovered at a luminosity $\sim$10$^{35}$ ergs
s$^{-1}$ in a 1993 {\it ASCA} observation (Torii et al.\ 1998; Gotthelf \&
Vasisht 1998), yet follow-up observations 3 and 6 years later found the flux 20
times dimmer (Vasisht et al.\ 2000).  No high-luminosity SGR-like bursts have
ever been seen from either of these sources.

One of the best studied AXPs, \axpb, has also shown signs of flux variability
(e.g.\ Oosterbroek et al.\ 1998).  Weekly monitoring with {\it RXTE} has
revealed two pulsed flux flares lasting several months (Gavriil \& Kaspi
2004).  Unlike the burst induced variability, the rises of these flares were
resolved lasting a few weeks.  Interestingly, two small bursts were detected
near the peak of the first flare (Gavriil et al.\ 2002), but none were seen at
any point during the much brighter and longer-lived second flare.  Imaging
X-ray observations have shown similar varibility in the phase-averaged
luminosity, albeit with much sparser sampling (Mereghetti et al.\ 2004).  As
with \axpf\ and \axpg, the ``baseline'' luminosity of \axpb\ is low ($\sim$6
$\times$ 10$^{33}$ ergs s$^{-1}$) relative to the average luminosity of the
AXP class.

The realization that SGRs and AXPs can enter low states with luminosities of
order 10$^{33}-10^{34}$ ergs s$^{-1}$ for extended periods of time has 
important implications on the total number density of magnetar candidates in
our Galaxy (\S14.8).  Their duty cycle as bright X-ray sources is presently
unknown, especially as a function of age.  There is still much to be
learned about the similarities and differences between magnetar candidates
in their dim states, and other low-luminosity X-ray sources such as 
Isolated Neutron Stars, Compact Central Objects and high-field radio
pulsars (\S7).  It is possible that some of these other sources occasionally
become X-ray bright like the AXPs.

\subsection{X-ray Spectra}

The X-ray spectra of SGRs and AXPs (0.5$-$10 keV) are usually well fit by a
two-component model, a blackbody plus a power law, modified by interstellar
absorption (Table~\ref{tab:spectra}).  The soft blackbody component is not 
required in a few sources, but these tend to be dim and/or heavily absorbed
(e.g.\ \sgrd).  During quiescence (i.e., outside of bursting activity), the
blackbody temperature does not vary greatly between different members of the
class (Marsden \& White 2001) or with time for individual sources (e.g.\
Oosterbroek et al.\ 1998).  On the other hand, the non-thermal component does
show significant variations between different sources (Marsden \& White 2001)
and with time in a few cases (e.g.\ Woods et al.\ 2004).


\begin{table}[!h]
\begin{minipage}{1.0\textwidth}
\caption{X-ray spectral properties of the SGRs and AXPs. \label{tab:spectra}}
\label{tbl:spectrum}
\vspace{10pt}

\begin{tabular}{cccccc}
\hline
Source$^{a}$                &
N$_{\rm H}$                 &
Blackbody                   &
Photon                      &  
Unabsorbed$^{b}$            &  
Luminosity$^{c}$            \\  
                            &
                            &
Temperature                 &
Index                       &
Flux                        &
                            \\
                            &
10$^{22}$                   &
                            &
                            &
10$^{-11}$                  &
10$^{35}$                   \\
                            &
(cm$^{-2}$)                 &
(keV)                       &
                            &
(ergs cm$^{-2}$ s$^{-1}$)   &
(ergs s$^{-1}$)             \\

\hline


SGR~0526$-$66    &  0.55  &  0.53 &  3.1       & 0.087        & 2.6        \\
SGR~1627$-$41    &   9.0  &  $-$  &  2.9       & 0.027$-$0.67 & 0.04$-$1.0 \\
SGR~1806$-$20    &   6.3  &  $-$  &  2.0       &  1.2$-$2.0   & 3.2$-$5.4  \\
SGR~1900$+$14    &   2.6  &  0.43 &  1.0$-$2.5 &  0.75$-$1.3  & 2.0$-$3.5  \\

\hline

CXOU~010043.1$-$721134 &  0.14  &  0.41  &  $-$      &  0.010      &  0.39         \\
4U~0142$+$61           &  0.91  &  0.46  &  3.4      &  8.3        &  0.72         \\
1E~1048.1$-$5937       &   1.0  &  0.63  &  2.9      &  0.41$-$2.3 &  0.053$-$0.25 \\
1RXS~J170849$-$400910  &   1.4  &  0.44  &  2.4      &  6.4        &  1.9          \\
XTE~J1810$-$197$^{d}$  &   1.1  &  0.67  &  3.7      &  0.01$-$2.2 &  0.01$-$2.6   \\
1E~1841$-$045          &   2.5  &  0.44  &  2.0      &  1.9        &  1.1          \\
AX~J1845$-$0258        &     9  &  $-$   &  4.6      & 0.04$-$1.0  &  0.05$-$1.2   \\
1E~2259$+$586          &   1.1  &  0.41  & 3.6$-$4.2 & 1.6$-$5.5   &  0.17$-$0.59  \\ 

\hline

\end{tabular}

\begin{flushleft}
{\small 
$a$ -- Spectral values given for quiescent state only (i.e.\ periods with no 
{\it detected} burst activity) \\
$b$ -- All fluxes and luminosities integrated over 2.0$-$10.0 keV \\
$c$ -- Assumed distances given in Table 14.4 \\
$d$ -- Spectral parameters given were obtained during ``high'' state of source
following its discovery in 2003  \\

\vspace{11pt}

REFERENCES -- (SGR~0526) Kulkarni et al.\ 2003; (SGR~1627) Kouveliotou et al.\
2003; (SGR~1806) Mereghetti et al.\ 2000; (SGR~1900) Woods et al.\ 2001;
(CXO~0100) Lamb et al.\ 2002; (4U~0142) Patel et al.\ 2003; (1E~1048)
Mereghetti et al.\ 2004; (RXS~1708) Rea et al.\ 2003; (XTE~1810) Gotthelf et
al.\ 2004; (1E~1841) Morii et al.\ 2003; (AX~1844) Gotthelf \& Vasisht 1998,
Torii et al.\ 1998, Vasisht et al.\ 2000; (1E~2259) Woods et al.\ 2004

}
\end{flushleft}


\end{minipage}
\end{table}

\begin{figure}[!htb]

\centerline{
\psfig{file=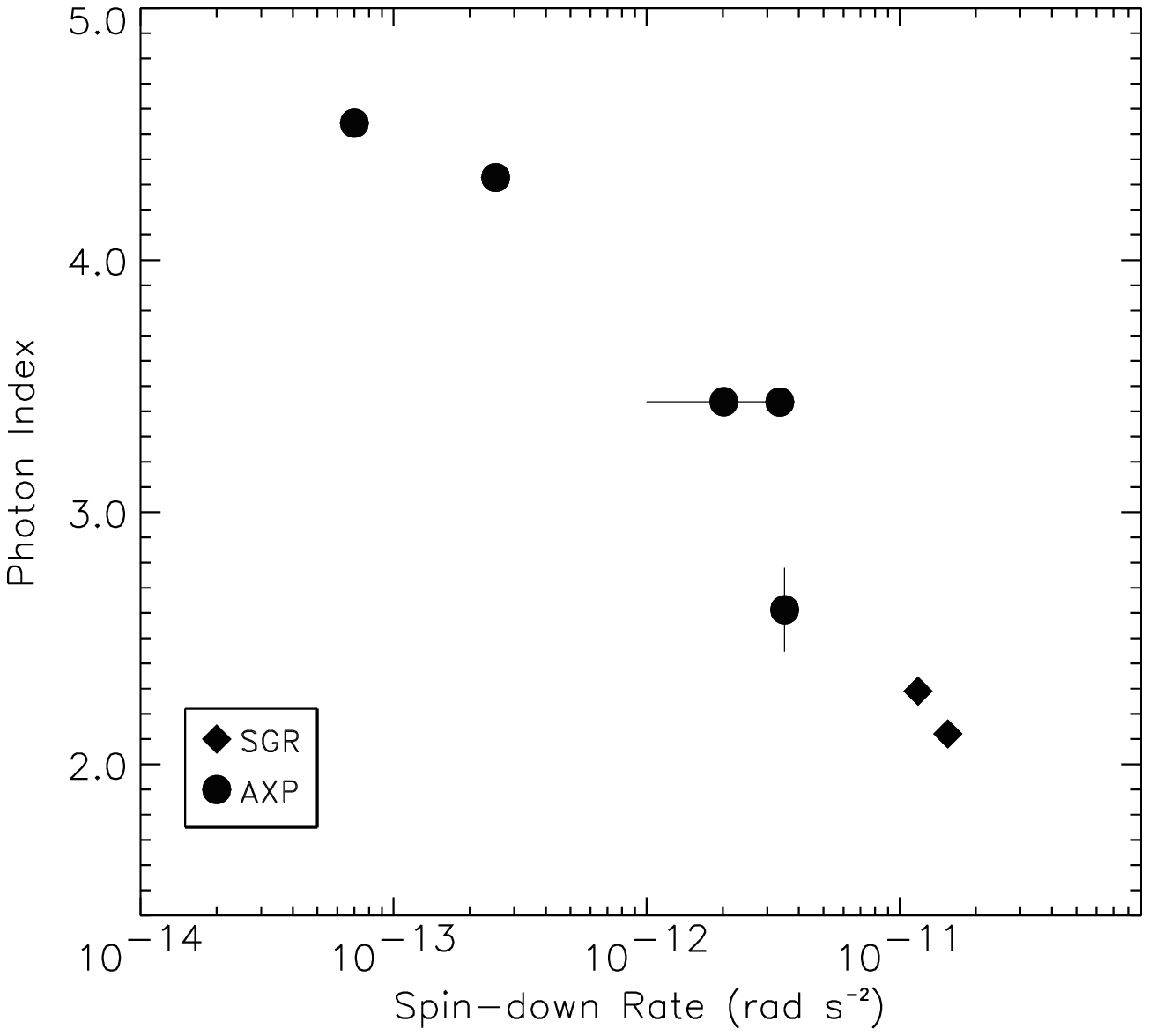,height=3.5in}}
\vspace{-0.1in}

\caption{The variation of the single power law photon index versus  spin-down
rate $|\dot{\Omega}|$ for each SGR and AXP.  The results  for objects with
more than one observation have been averaged. The photon  index decreases
(spectral hardness increases) with increasing spin-down  rate. Figure from
Marsden \& White (2001). \label{fig:xray_spec}}

\end{figure}

The first systematic study of the X-ray spectra of SGRs and AXPs was performed
by Marsden \& White (2001), who found that the spectral hardness of the
persistent X-ray counterparts of these sources formed a continuum and was
positively correlated with the spin-down rate of the pulsar
(Figure~\ref{fig:xray_spec}). The varying hardness of the X-ray spectrum with
spin-down rate was  linked to the non-thermal component of the spectrum.


Until the launch of {\it Integral}, very little was known about the X-ray
spectra of magnetar candidates above $\sim$10 keV due to the limitations of
past instrumentation.  Currently, there have been reported {\it Integral}
detections of persistent hard X-ray emission above $\sim$15 keV from four
magnetar candidates (\axpe\ [Molkov et al.\ 2004; Bassani et al.\ 2004], \axpd\
[Revnivtsev et al.\ 2004], \sgra\ [Bird et al.\ 2004], and \axpc\ [den Hartog
et al.\ 2004]).  Kuiper, Hermsen \& Mendez (2004) were the first to show that
the hard X-rays detected from \axpe\ do, in fact, originate with the AXP when
they detected pulsed emission with {\it RXTE} HEXTE.  Interestingly, the pulsed
spectrum follows a power law with a photon index $-$1.0 up to at least 100
keV.  Knowledge of the photon distribution with energy above 15 keV is crucial 
to determining the underlying emission mechanism (see \S14.7.3).

There are no definite detections of spectral features in the {\it persistent}
X-ray emission of magnetar candidates.  Grating spectra from {\it Chandra} and
{\it XMM-Newton} have yielded strong upper limits ($<$30 eV) on narrow line
features for \axpc\ (Juett et al.\ 2002) and \axpa\ (Woods et al.\ 2004) in the
energy range 0.5$-$5 keV. The only reported detection of a spectral feature
comes from a {\it BeppoSAX} spectrum of \axpd\ where Rea et al.\ (2003) find a
$\sim$4$\sigma$ absorption line at $\sim$8 keV.  The absence of spectral
features at X-ray energies where proton-cyclotron resonances would occur  in
magnetar-strength fields places strong constraints on models of the
transmission of heat through the surface, and of surface heating.

\subsection{Pulse Profiles and Pulsed Fractions}

\begin{figure}[!htb]

\centerline{
\psfig{file=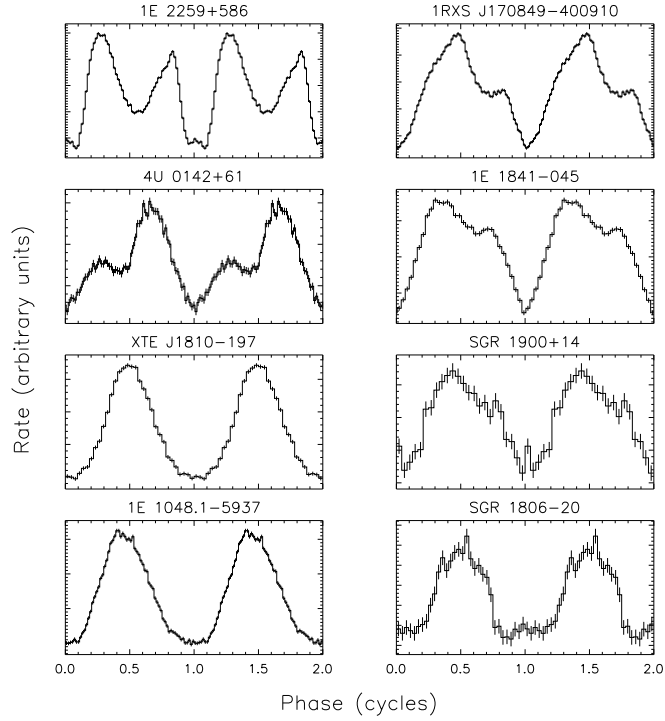,height=3.8in}}
\vspace{-0.1in}

\caption{The folded pulse profiles of eight different magnetar candidates.  The
sources are ranked according to inferred dipole magnetic field strength. 
Magnetic field increases from top to bottom and left to right.  All profiles
are of X-rays between 2 and 10 keV as observed with the {\it RXTE} PCA.  Note
that the folded profile of \sgrc\ is from after the August 27 flare.  AXP pulse
profiles courtesy of V.M. Kaspi and F.P. Gavriil. \label{fig:profiles}}

\end{figure}

The X-ray pulse profiles of magnetar candidates range from simple 
sinusoids to more complex profiles showing (typically) two maxima  per
cycle (Figure~\ref{fig:profiles}).  The observed pulse morphologies of the AXPs
are consistent with either one or two hot spots on the surface of a  neutron
star (\"Ozel 2002), but the spectrally harder SGRs sometimes have more
complicated pulse profiles.  In contrast with most accreting X-ray pulsars, the
pulse profile of a SGR or AXP often has a weak dependence on photon
energy.



The pulse profiles shown in Figure~\ref{fig:profiles} are ranked in order of
increasing spin-down rate (from top to bottom and left to right).  No strong
trend is apparent; but note that the pulse profile of \sgrc\ was much more 
complex before the August 27 flare.  The evolution of the pulse morphology in
\sgrc\ and other magnetar candidates is discussed in \S14.5.3.

The root-mean-square (rms) X-ray pulsed fractions of magnetar candidates range
from 4 to 60\% (Table~\ref{tab:timing}).  Note that in the literature, both
peak-to-peak and rms are reported, and the rms values are always less than the
peak-to-peak values for any given AXP/SGR pulse profile.  Similar to the pulse
shape, the pulsed fractions of the AXPs show little or no change with photon
energy (0.5$-$10 keV).  Since the relative contribution of the blackbody
spectral component to the total photon flux changes from 0\% to as much as
$\sim$70\% over this bandpass, \"Ozel, Psaltis \& Kaspi (2001) argued that the
two spectral components of the AXPs must by highly correlated or caused by the
same physical process.  The pulsed fraction also places strong constraints
on models in which the 2-10 keV emission of the AXPs is purely due to
cooling emission through the surface of the star (\"Ozel et al. 2001).  
It should be kept in mind that emission due to surface heating can be much
more strongly beamed (e.g.\ Basko \& Sunyaev 1974); and that cyclotron 
scattering by persistent electric currents can strongly modify the observed
pulse profile in active magnetars (Thompson et al.\ 2002).

\section{Timing Behavior}

\begin{figure}[!htb]

\centerline{
\psfig{file=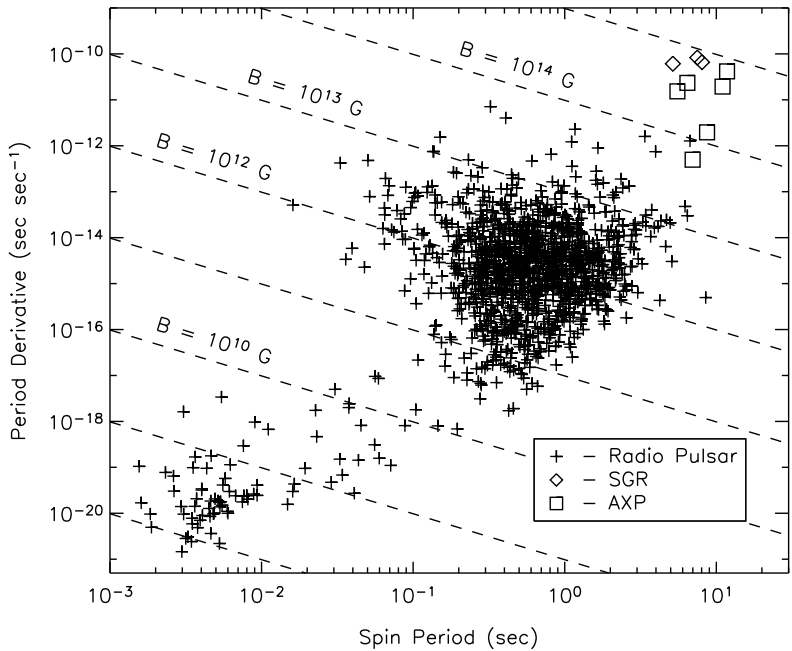,height=3.7in}}

\caption{Period versus period derivative for radio pulsars (plus signs),
Anomalous X-ray Pulsars (squares), and Soft Gamma Repeaters (diamonds). 
Contours of constant inferred magnetic field strength are drawn as diagonal
dashed lines.  Radio pulsar data courtesy of the ATNF Pulsar Group$^{\ddag}$. 
AXP and SGR timing data are given in Table 14.2. \label{fig:pvspdot}}

\end{figure}

The spin periods of the SGRs and AXPs are clustered between 5 and 12 seconds, a
very narrow range compared with radio pulsars and accreting X-ray pulsars.  
These sources are all spinning down rapidly and persistently, with
fairly short characteristic ages $P/\dot P \sim 10^3-10^5$ yrs.   The magnitude
of the spin-down torque is consistent with magnetic dipole braking of an
isolated neutron star with a dipole field of  $\sim 10^{14}-10^{15}$ G
(Figure~\ref{fig:pvspdot}).  Although most of the characteristic ages are 
less than $10^4$ yrs, the
ages for individual sources should be treated with caution since the spin down
torque has been observed to vary by more than a factor $\sim 4$ in the
SGRs \sgra\ and \sgrc.
The pulse timing properties are summarized in Table~\ref{tab:timing}.


\begin{table}[!h]
\begin{minipage}{1.0\textwidth}
\begin{center}
\caption{Pulse timing properties of the SGRs and AXPs. \label{tab:timing}}
\label{tbl:timing}
\vspace{10pt}

\begin{tabular}{cccccc}
\hline
Source                      &
Period                      &
Period                      &
Magnetic                    &  
Spin down                    &
Pulsed                      \\  
                            &
                            &
Derivative                  &
Field$^{a}$                 &
Age$^{b}$                   &
Fraction$^{c}$              \\
                            &
(s)                         &
(10$^{-11}$ s s$^{-1}$)     &
(10$^{14}$ Gauss)           &
(10$^{3}$ years)            &
(\% rms)                    \\

\hline

SGR~0526$-$66    &  8.0   &  6.6       &  7.4   &  1.9  &  4.8  \\
SGR~1627$-$41    &  6.4?  &  $-$       &  $-$   &  $-$  & $<$10 \\
SGR~1806$-$20    &  7.5   &  8.3$-$47  &  7.8   &  1.4  &  7.7  \\
SGR~1900$+$14    &  5.2   &  6.1$-$20  &  5.7   &  1.3  & 10.9  \\

\hline

CXOU~010043.1$-$721134 &   8.0  &  $-$   &  $-$   &  $-$  & 10     \\
4U~0142$+$61           &   8.7  &  0.20  &  1.3   &  70   &  3.9   \\
1E~1048.1$-$5937       &   6.4  &  1.3$-$10   &  3.9   &  4.3  & 62.4   \\ 
1RXS~J170849$-$400910  &  11.0  &  1.9   &  4.7   &  9.0  & 20.5   \\
XTE~J1810$-$197        &   5.5  &  1.5   &  2.9   &  5.7  & 42.8   \\
1E~1841$-$045          &  11.8  &  4.2   &  7.1   &  4.5  & 13     \\
AX~J1844$-$0258        &   7.0  &  $-$   &  $-$   &  $-$  & 48     \\
1E~2259$+$586          &   7.0  &  0.048 &  0.60  &  220  & 23.4   \\

\hline

\end{tabular}

\begin{flushleft}
{\small 
$a$ -- $B_{\rm dipole} = 3.2 \times 10^{19} \sqrt{{\rm P}\dot{\rm P}}$ G 
(the mean surface dipole field)\\
$b$ -- Characteristic age of pulsar spinning down via magnetic braking
(P/2$\dot{\rm P}$) \\
$c$ -- $f_{\rm rms} = \sqrt{\frac{1}{N} \sum_{i=1}^{N} (r_i - r_{\rm avg})^2 - 
e_i^2}/r_{\rm avg}$, where $N$ = number of phase bins, $r_i$ is the count rate
(2$-$10 keV) in the $i^{\rm th}$ phase bin, and $e_i$ is the error in the  
rate \\

\vspace{11pt}

REFERENCES -- (SGR~0526) Kulkarni et al.\ 2003; (SGR~1627) Woods et al.\ 1999a;
(SGR~1806) Woods et al.\ 2002; (SGR~1900) Woods et al.\ 2002; (CXO~0100) Lamb
et al.\ 2003b; (4U~0142) Gavriil \& Kaspi 2002;  (1E~1048) Kaspi et al.\ 2001;
Gavriil \& Kaspi 2004; (RXS~170849) Gavriil \& Kaspi 2002; (XTE~1810) Ibrahim
et al.\ 2004; (1E~1841) Gotthelf et al.\ 2002; (AX~1844) Gotthelf \& Vasisht
1998, Torii et al.\ 1998; (1E~2259) Gavriil \& Kaspi 2002

}
\end{flushleft}

\end{center}

\end{minipage}
\end{table}

Long-term phase-coherent timing of SGRs and AXPs recently became feasible with
{\it RXTE}.  Currently several AXPs have continuous timing solutions, with some
dating back to 1998 (e.g.\ Gavriil \& Kaspi 2002).  In the case of two AXPs
(\axpb\ [Kaspi et al.\ 2001]; \axpf\ [Ibrahim et al.\ 2004]) and two SGRs
(\sgra\ and \sgrc\ [Woods et al.\ 2002])  phase-coherent timing is not always
possible but has been obtained over stretches of months to years.  Major
results of this timing effort have been the discoveries of three glitches from
two AXPs  and strong timing noise detected in both SGRs and the AXP \axpb.


\footnotetext[2]{\tt http://www.atnf.csiro.au/research/pulsar/psrcat/}

The occurrence of large glitches is a natural consequence of the magnetar model
(Thompson \& Duncan 1996) and was even suspected to be the primary source of
timing noise in various models of the AXPs (Usov 1994;  Heyl \& Hernquist
1999).   In 1999, the first glitch from an AXP was observed by {\it RXTE} in
the source \axpd\ (Kaspi, Lackey \& Chakrabarty 2000).  Since  that time,
another glitch was detected from the same source (Kaspi \&  Gavriil 2003;
Dall'Osso et al.\ 2003) and one glitch was observed in  \axpa\ coincident with
a burst active episode (Kaspi et al.\ 2003;  Woods et al.\ 2004).  No glitches
have been directly observed in the  SGRs, although \sgrc\ has shown evidence
for rapid spin-{\it down} at  the time of the August 27 flare (see \S14.5.4). 
The magnitude of the angular momentum exchange within the star that one infers
for the AXP  glitches is more characteristic of Crab-type pulsar glitches than
the larger Vela-type glitches, but there are some dissimilarities with radio
pulsars glitch behavior (Table~\ref{tab:glitches}).


\begin{table}[!h]
\begin{minipage}{1.0\textwidth}
\begin{center}
\caption{Properties of the three glitches observed in two AXPs. \label{tab:glitches}} 
\label{tbl:glitches}
\vspace{10pt}
\begin{tabular}{lccc} \hline 

Source   & 1RXS~J1708$-$40 & 1RXS~J1708$-$40  &  1E~2259$+$586    \\

$\Delta\nu/\nu^a$ & $5.5 \times 10^{-7}$ & $1.4 \times 10^{-7}$ & $3.7 \times 10^{-6}$ \\
$\Delta \nu_{g}/\nu$  & ... & ... & $> 6.1 \times 10^{-6}$ \\
$\tau_{g}$ (days) & ... & ... & 14 \\
$\Delta \nu_d$  & ... & $4.1 \times 10^{-6}$ & $\sim\Delta \nu_{g}/\nu$ \\
$\tau_d$ (days) & ... & 50 & 16 \\
$\Delta \dot{\nu}/\dot{\nu}$  & $-0.010$  & $<|0.001|$ &  $+0.022$ \\
$t_{glitch}$ (MJD TDB) & 51444.6 & 52014.2 & 52443.1 \\

\hline
\end{tabular}

\begin{flushleft}

{\small 
\noindent${a}$ -- Frequency denoted by $\nu$ and frequency derivative by
$\dot{\nu}$.  The subscript $g$ indicates the frequency growth terms and $d$
indicates decay terms. \\

\vspace{11pt}

REFERENCES -- (1RXS~1708) Kaspi \& Gavriil 2003; Dall'Osso et al.\ 2003;
(1E~2259) Woods et al.\ 2004
}

\end{flushleft}

\end{center}

\end{minipage}
\end{table}

The glitch observed from \axpa\ was especially interesting in that
it coincided with a SGR-like outburst, and also with changes in 
the X-ray flux and pulse profile that persisted for months (see \S14.5.3).
There is evidence for a very long-term component of the 
post-glitch frequency recovery (consistent with a persistent change
in torque) in this glitch, as well as in one of the glitches of \axpd.
Overall one observes a great diversity in behavior even within this small
sample of glitch events, and more extended monitoring is required
to unravel the relationship between glitch behavior and burst activity.

In addition to rapid spin down, all SGRs and AXPs have shown significant 
timing noise:  an irregular drift of the spin frequency superposed on the 
secular spin down trend.  Recent timing solutions have shown that most 
of this noise is not caused by resolved glitches.  The existence of 
timing noise was first noted for the AXPs \axpa\ (e.g.\ Baykal \& Swank 1996)
and \axpb\ (e.g.\ Oosterbroek et al.\ 1998).  Its strength was estimated 
in four AXPs by Heyl \& Hernquist (1999), and was found to be marginally 
consistent with an extrapolation of the correlation between timing 
noise strength and braking torque observed in radio pulsars 
(e.g.\ Arzoumanian et al.\ 1994).  

In the {\it RXTE} era, the first direct detection of timing noise was made in
\sgra\ (Woods et al.\ 2000).  This SGR is one of the ``noisiest'' rotators
among SGRs and AXPs and has shown a long-term persistent change in torque,
along with large stochastic offsets in the X-ray pulse phase on timescales as
short as $\sim 10^4$ s (Woods et al.\ 2002).  Although the strength of the
timing noise  is consistent with some of the ``quieter'' accreting X-ray
pulsars, the shape of the torque power spectrum is more similar to that
observed in radio pulsars.  Overall, the timing analysis of the AXPs (Kaspi et
al.\ 2001; Gavriil \& Kaspi 2002) and one other SGR (Woods et al.\ 2002) has
revealed a broad range of torque variability, with some evidence for a
correlation between the strength of the timing noise and the spin-down rate.

\section{Burst-Induced Variability}

It has become evident that burst activity in the SGRs can have a 
persistent effect on the underlying X-ray source.  During the 1998 burst
activation of \sgrc, the X-ray counterpart became brighter, its energy spectrum
was altered, and the pulse shape changed dramatically.  Furthermore, the X-ray
counterpart to \sgrd\ has become progressively dimmer since the one recorded
outburst from this SGR in 1998.  Finally, the AXP \axpa\ showed a broad array
of spectral and temporal changes coincident with its 2002 outburst.  We now
present some details of the burst-induced variability that is observed 
in magnetar candidates.

\subsection{X-ray Afterglows and AXP Outbursts}

Extended X-ray afterglow has been detected
following four separate bursts from \sgrc.  The first such detection
followed the giant flare of 1998 August 27 (Woods et al.\ 2001). 
One half hour following the flare, the persistent X-ray flux from
\sgrc\ remained $\sim$700 times brighter than the pre-flare level.  The
X-ray flux decayed over the next 40 days approximately as a
power law in time ($F \propto t^{-\alpha}$ with an exponent $\alpha =
0.71$).  The blackbody component of the X-ray spectrum was hotter 
($kT =$ 0.94) one day into the afterglow phase
than it was before the burst ($kT =$ 0.5 keV); but eighteen days 
later the power-law component of the spectrum was again dominant.



\begin{figure}[!htb]
\centerline{
\psfig{file=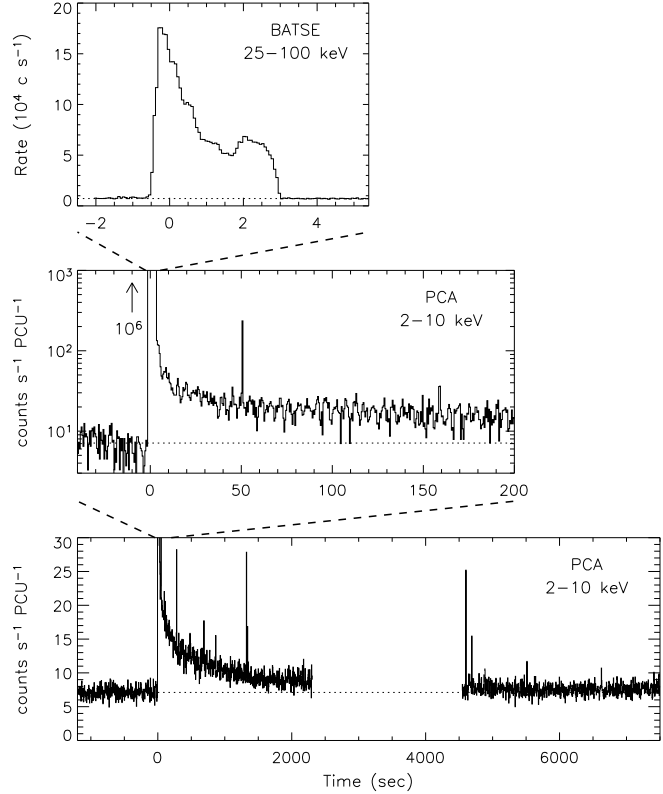,height=4.2in}}

\caption{The time history of the August 29$^{\rm th}$ burst and its afterglow. 
{\it Top} - The BATSE light curve showing the sharp rise and fall of the
burst.  {\it Middle} - The PCA light curve including the burst (off-scale) and
the early portions of the afterglow (T$+$3 s).  There is a sharp discontinuity
in the energy spectrum when the high-luminosity burst emission terminates at 3
s (Ibrahim et al.\ 2001) indicating the transition from the burst to the
afterglow.  Note the clear 5.16 s pulsations in the light curve.  {\it Bottom}
- The PCA light curve over a longer time interval showing the gradual decay of
the afterglow.  The spectral evolution during the afterglow is presented in
Ibrahim et al.\ (2001) and Lenters et al.\ (2003).  The horizontal dotted lines
in all panels represent the background level. \label{fig:aug29}}

\end{figure}

Afterglows have also been detected from \sgrc\ following bursts  on 1998 August
29 (Ibrahim et al.\ 2001; Lenters et al.\ 2003 [Figure~\ref{fig:aug29}]), 2001
April 18 (Feroci et al.\ 2003), and 2001 April 28 (Lenters et al.\ 2003).  A 
power-law decay is also seen in these cases, with a return to the pre-burst
flux level between 10$^{4}$ and 10$^{6}$ s following the burst.   Enhanced
thermal emission is typical, with  temperatures as high as $\sim 4$ keV
(corresponding to a hot spot covering $\sim 1$ percent of the neutron star
surface).   In fact, the afterglow of the 2001 April 28 burst involved only
enhanced thermal emission.  Within this small sample of afterglows from \sgrc,
the $2-10$ keV afterglow energy is about 2\% of the $25-100$ keV burst energy
(Lenters et al.\ 2003).  

\begin{figure}[!htb]
\centerline{
\psfig{file=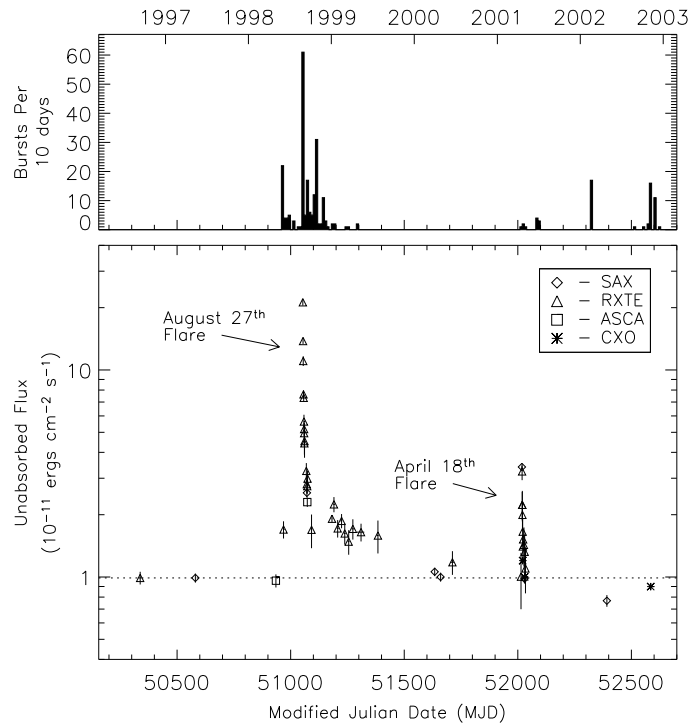,height=3.8in}}

\caption{{\it Top panel} -- Burst rate history of SGR~1900$+$14 as observed
with BATSE and the IPN.  {\it Bottom panel} -- Persistent/Pulsed flux history
of SGR~1900$+$14 covering 5.5 years.  The vertical scale is unabsorbed 2$-$10
keV flux.  The pulsed fraction is assumed constant to convert pulsed flux to
phase-averaged flux (see Woods et al.\ 2001 for details). The dotted line marks
the nominal quiescent flux level of this SGR.  Note the strong correlation
between the burst activity and the flux enhancements.
\label{fig:sgr1900_fhist}}

\end{figure}

Resolved observations of individual SGR burst afterglows are still rare, 
because they require pointed X-ray observations coincident with a burst
(or very soon thereafter).  It is much easier to observe the 
collective effect of SGR burst activity on the persistent X-ray flux,
as is seen in the case of \sgrc\ (Figure~\ref{fig:sgr1900_fhist}).  



The detection of X-ray bursts (Kaspi et al.\ 2003) from the AXP \axpa\ was the
fortunate result of a long-term monitoring campaign by {\it RXTE}. The X-ray
flux of this source increased by at least a factor $\sim$20 on 2002 June 18
(Woods et al.\ 2004), during which more than 80 SGR-like bursts were emitted
(Gavriil et al.\ 2004). This first component of the flux decay was spectrally
hard, contained all of the observed burst activity, and involved only $\sim 1$
percent of the neutron star surface.  It decayed within $\sim$1 day, and was
followed by a much more gradual flux decay over the following year. This more
extended X-ray brightening involved a significant fraction of  the warm stellar
surface, but only a modest spectral hardening.   No bright burst (similar to
the intermediate bursts of the SGRs) appears to have preceded this activity.

The longer term flux variability of the SGR and AXP sources is still unclear.
Some sources (such as the AXPs \axpa, \axpb, and \axpc) have remained X-ray 
bright for two-three decades.  A previous X-ray brightening of \axpa\ detected
10 years
earlier by Ginga (Iwasawa, Koyama \& Halpern 1992) allows one to deduce that at
least $\sim$10\% of the X-ray output of this source is released in transient
events.   A steady decrease in X-ray flux was also observed in \sgrd\ after its
outburst in June/July 1998,  but this was followed by a sharper drop a few
years later  (Kouveliotou et al.\ 2003).   The X-ray flux of this source has
appeared to level off at a value ($\sim$4$\times 10^{33}$ ergs s$^{-1}$)
consistent with the low state levels seen in  at least two other sources.  It
is possible that magnetar candidates become increasingly intermittent X-ray
sources as they age; alternatively, some intermittent sources may have weaker
magnetic fields.  There is, nonetheless, clear evidence for both short-term and
long-term  flux variability associated with X-ray outbursts.

\subsection{Transient Counterparts at other Wavelengths}

The only recorded radio emission from a magnetar candidate was associated with
the 1998 August 27 giant flare of \sgrc\ (Frail, Kulkarni \& Bloom 1999).  No
radio detection was made before or since; but a faint transient persisted for 2
weeks following the X-ray flare with a spectral index of $-$0.74$\pm$0.15.  No
similar radio detections have been made following other SGR outbursts (all of
which were much less energetic than the August 27 flare and probably involved a
much weaker particle outflow).

Optical and/or infrared (IR) counterparts have been discovered in four (possibly
five) magnetar candidates (see \S14.6.4).  The IR flux of the AXP \axpa\
increased  following its 2002 June outburst (Kaspi et al.\ 2003).  A week after
the  X-ray burst activity, the $K$-band flux was 3.4 times higher than measured
two years earlier;  it had returned near its pre-outburst level after $\sim$40
days (Israel et al.\ 2003a).

Automated telescopes such as ROTSE have observed SGRs during burst active
periods (Akerlof et al.\ 2000).  
However, the visual extinction toward the SGRs observed is extremely high and
so the acquired limits are not constraining.   Followup observations of SGR
bursts with robotic IR cameras with fast photometric capabilities now coming
on-line such as BLANK, could provide interesting constraints on the burst
mechanism.

\subsection{Changes in X-ray Pulse Shape and Pulsed Fraction}

\begin{figure}[!htb]
\centerline{
\psfig{file=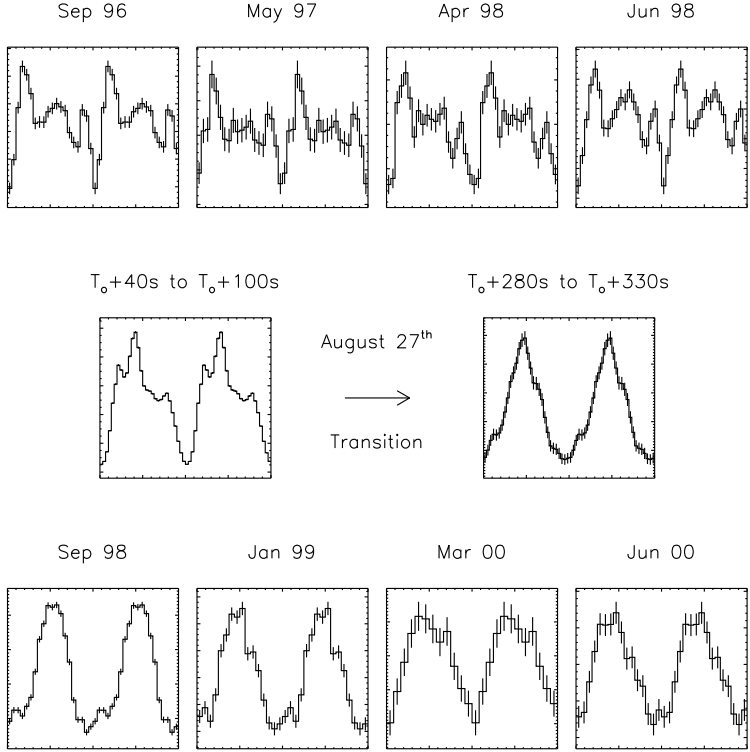,height=4.0in}}

\caption{Evolution of the pulse profile of SGR~1900$+$14 from 1996 September
through 2000 April (Woods et al.\ 2001).  All panels display two pulse cycles
and the vertical axes are count rates with arbitrary units.  The two middle
panels were selected from {\it Ulysses} data (25$-$150 keV) of the August
27$^{\rm th}$ flare.  Times over which the {\it Ulysses} data were folded are
given relative to the onset of the flare (T$_{\rm o}$).  The top and bottom
rows are integrated over the energy range 2$-$10 keV.  From top-to-bottom,
left-to-right, the data were recorded with the {\it RXTE}, {\it BeppoSAX}, {\it
ASCA}, {\it RXTE}, {\it RXTE}, {\it RXTE}, {\it BeppoSAX}, and {\it RXTE}.
\label{fig:profile_1900}}

\end{figure}

Changes in the X-ray pulse profile have been observed in magnetar candidates
during periods of intense burst activity.  The most profound changes in pulse
properties have been observed in \sgrc.  In particular, at the time of the
giant flare of 1998 August 27, the pulse profile of the persistent emission
changed dramatically from a complex, multi-peaked morphology to a simple,
nearly sinusoidal morphology (Figure~\ref{fig:profile_1900}).  The change in
pulse shape has persisted even years after the
post-flare afterglow faded away ({G\"o\u{g}\"u\c{s}} et al.\ 2002).   The
effectively permanent change in pulse shape observed in \sgrc\ argues for a
magnetic field reconfiguration at the time of the giant flare (Woods et al.\
2001).  A similar  simplification of the pulse shape was observed -- at a much
higher flux  level -- over the last few minutes of the August 27 flare
(Figure~\ref{fig:profile_1900}).   This change in pulse profile was smooth
and gradual; indeed, the flux decline was consistent with cooling of a
magnetically confined plasma  (Feroci et al. 2001).  Thus, the magnetic field
reconfiguration  may well have been concentrated in the initial impulsive phase
of the flare.



A significant change in pulse shape was also observed in \sgra\ between 
1996 and 2001 ({G\"o\u{g}\"u\c{s}} et al.\ 2002).   The change was
not as dramatic as in \sgrc; but the source also emitted far less
energy in X-ray bursts than was released in the August 27 flare.
More subtle changes in pulse shape were also observed in \sgrc\ between
1999 and 2001, when it underwent intermittent burst activity.
During the 2002 June 18 outburst of \axpa, the pulse profile evolved rapidly
showing large changes in the relative amplitudes of the two peaks (Kaspi et
al.\ 2003; Woods et al.\ 2004).  Some residual change has persisted at least
until one year after the burst activity. 
In sum, the connection between X-ray outbursts and pulse shape changes
can be subtle, and frequent monitoring of magnetar candidates
will be required to understand it better.

Large changes in pulsed fraction have been observed in magnetar candidates
during periods of burst activity.  During burst afterglows of \sgrc, the  rms
(2$-$10 keV) pulsed fraction has been observed  to rise to $\sim$20$-$30\% from
its quiescent value of $\sim$11\% (Lenters et al.\ 2003).  Intriguingly, the 
enhanced pulsations remain in phase with the pre-burst pulsations  in at least
two bursts.  This indicates a direct correlation between the source of the
pulsed X-ray emission and the active burst region.

A pulsed fraction change was also observed in \axpa\ during its 2002 June
outburst (Woods et al.\ 2004).  During the {\it RXTE} observation where the
burst activity was observed, the pulsed fraction (2$-$10 keV) actually {\it
decreased} to $\sim$15\% from the pre-outburst level of 23.4\%.  The pulsed
fraction recovered to $\sim$23.4\% within $\sim$6 days of the outburst, much
more rapidly than the pulse shape recovered.

\subsection{Connection with Timing Anomalies}

X-ray burst activity appears to have a variety of effects on the  spin behavior
of magnetar candidates.  A comparison of the spin evolution of \sgrc\ before
and after the August 27 flare showed that the source underwent a transient
spin down within an 80 day window that bracketed the flare (Woods et al.\
1999c).  The decrease in spin frequency, by one part in $10^4$, was opposite in
sign to pulsar glitches.  A comparison of the pulse timing during and after the
flare indicate that the change in frequency occurred within several hours of
the flare (Palmer 2002). The amplitude of the spin down is consistent with an
enhanced magnetic torque due to a relativistic outflow of particles (Thompson
\& Blaes 1998; Frail et al.\ 1999; Thompson et al. 2000) if the dipole field of
the star is $\sim$10$^{15}$ G.

Only the giant flare from \sgrc\ has shown direct evidence for burst-induced
spin down.  This SGR suffered another timing anomaly at the
time of the 2001 April 18 flare, but sparse data coverage did not allow for an
unambiguous determination of its nature (Woods et al.\ 2003).  More than 5
years of timing and burst data for each SGR have revealed longer-term increases
in braking torque, of similar magnitudes, that are not however synchronized
with burst activity.  (Indeed the output in X-ray bursts from \sgra\ was much
smaller than from \sgrc\ over this time interval.) There is, nonetheless, some
tentative evidence for a causal relation between burst activity and torque
variability, in that 
the most burst-active magnetar candidates
are also those which  show the strongest timing noise.

A timing anomaly of a different type,  the glitch of \axpa\ (\S14.4),
coincided with the 2002 June X-ray outburst (Kaspi et al.\
2003; Woods et al.\ 2004).  Since the beginning of
the X-ray activity was not observed, it was not possible
to determine whether it preceded, overlapped, or followed the
onset of the glitch.  Nonetheless, the change in rotational
energy associated with the glitch was much smaller than the 
energy released in the X-ray transient, suggesting that the
trigger involved some other agent (e.g., the release of magnetic stresses).  

In the case of the SGRs, glitches of the magnitude seen in
\axpa\ cannot generally be excluded:  the timing ephemerides preceding
X-ray outbursts are less accurate for these sources.
However, X-ray transients associated with glitches of other AXP sources
are easier to constrain.  No burst emission was seen near the time of either of
the two glitches observed in \axpd\ (Kaspi \& Gavriil 2003). 
A $\sim1$ day hard-spectrum transient, such as was observed from \axpa, 
could easily have been missed.  However, the \axpa\ outburst also showed a
sustained X-ray afterglow lasting months and a pulse profile change for a
somewhat shorter time interval (Woods et al.\ 2004).  The regular
monitoring of \axpd\ would have been sensitive to changes of this magnitude.
Continued phase-coherent timing of these objects is needed to determine
the extent to which glitches are accompanied by burst activity.

\section{Locations, SNR Associations, and Counterparts}

A multi-wavelength approach has always proven fruitful for
understanding enigmatic astrophysical objects.  The SGRs and AXPs
are no exception to this rule.  
Sub-arc-second determinations of their X-ray locations 
(Table~\ref{tab:locations})
have allowed follow-up observations at radio and optical/IR wavelengths.
(In the case of the SGRs, these represent a considerable refinement
over previous triangulation of burst emissions using the Interplanetary
Network.)
We now discuss the results of this collective effort to study
the magnetar candidates.



\begin{table}[!h]
\begin{minipage}{1.0\textwidth}
\begin{center}
\caption{The X-ray positions, reported associations, and the inferred distances
of the SGRs and AXPs. \label{tab:locations}}
\label{tbl:radio}
\vspace{10pt}

\begin{tabular}{cccccc}
\hline
Source                      &
Right                       &
Declination                 &
Associated                  &  
Distance                    &
Galactic                    \\  
                            &
Ascension$^{b}$             &
                            &
SNR/Cluster                 &
                            &
Scale Height                \\
                            &
(J2000)                     &
(J2000)                     &
                            &
(kpc)                       &
(pc)                        \\

\hline


SGR~0526$-$66    &   05$^{\rm h}$ 26$^{\rm m}$ 00.89$^{\rm s}$  &
	$- 66^{\circ}$ 04$^{\prime}$ 36.3$^{\prime \prime}$  &
        N49/cluster?     &  50      &  n/a  \\
SGR~1627$-$41    &  16$^{\rm h}$ 35$^{\rm m}$ 51.84$^{\rm s}$  &
	$- 47^{\circ}$ 35$^{\prime}$ 23.3$^{\prime \prime}$     &
        ...      &      11  &  $-$21  \\
SGR~1801$-$23$^{b}$ &  18$^{\rm h}$ 00$^{\rm m}$ 59$^{\rm s}$  &
	$- 22^{\circ}$ 56$^{\prime}$ 50$^{\prime \prime}$     &
        ...      & $\sim$10 &  ...  \\
SGR~1806$-$20    &  18$^{\rm h}$ 08$^{\rm m}$ 39.32$^{\rm s}$  &
	$- 20^{\circ}$ 24$^{\prime}$ 39.5$^{\prime \prime}$     &
        cluster  &      15  &  $-$63   \\
SGR~1900$+$14    &  19$^{\rm h}$ 07$^{\rm m}$ 14.33$^{\rm s}$  &
	$+ 09^{\circ}$ 19$^{\prime}$ 20.1$^{\prime \prime}$     &
        cluster  &      15  &  $+$200  \\

\hline

CXOU~010043.1$-$721134 &   01$^{\rm h}$ 00$^{\rm m}$ 43.14$^{\rm s}$  &
	$- 72^{\circ}$ 11$^{\prime}$ 33.8$^{\prime \prime}$     &
        ...     &  57      &  n/a  \\
4U~0142$+$61           &   01$^{\rm h}$ 46$^{\rm m}$  22.42$^{\rm s}$  &
	$+ 61^{\circ}$ 45$^{\prime}$ 02.8$^{\prime \prime}$     &
        ...      &      3   &  $-$20  \\
1E~1048.1$-$5937       &   10$^{\rm h}$ 50$^{\rm m}$ 07.14$^{\rm s}$  &
	$- 59^{\circ}$ 53$^{\prime}$ 21.4$^{\prime \prime}$     &
        ...      &      3  &  $-$27  \\
1RXS~J170849$-$400910  &   17$^{\rm h}$ 08$^{\rm m}$ 46.87$^{\rm s}$  &
	$- 40^{\circ}$ 08$^{\prime}$ 52.4$^{\prime \prime}$     &
        ...      &      5  &  $+$3  \\
XTE~J1810$-$197        &   18$^{\rm h}$ 09$^{\rm m}$ 51.08$^{\rm s}$  &
	$- 19^{\circ}$ 43$^{\prime}$ 51.7$^{\prime \prime}$     &
        ...      & $\sim$10 &  ...  \\
1E~1841$-$045          &   18$^{\rm h}$ 41$^{\rm m}$ 19.34$^{\rm s}$  &
	$- 04^{\circ}$ 56$^{\prime}$ 11.2$^{\prime \prime}$     &
        G27.4$+$0.0 &   7  &  $-$1  \\
AX~J1844$-$0258$^{c}$  &   18$^{\rm h}$ 44$^{\rm m}$ 53$^{\rm s}$  &
	$- 02^{\circ}$ 56$^{\prime}$ 40$^{\prime \prime}$       &
        G29.6$+$0.1 &  $\sim$10  &  $+$20  \\
1E~2259$+$586          &   23$^{\rm h}$ 01$^{\rm m}$ 08.30$^{\rm s}$  &
	$+ 58^{\circ}$ 52$^{\prime}$ 44.5$^{\prime \prime}$     &
        G109.1$-$1.0 &  3  &  $-$52  \\

\hline

\end{tabular}

\begin{flushleft}
{\small 
$a$ -- All positions accurate to $<$1$^{\prime \prime}$ unless otherwise noted \\
$b$ -- Only a very crude IPN location ($\sim$80 arcmin$^2$ area) exists for this 
SGR \\
$c$ -- The positional accuracy for this AXP is a 20$^{\prime \prime}$ radius
circle \\

\vspace{11pt}

REFERENCES -- (SGR~0526) Kulkarni et al.\ 2003; Klose et al.\ 2004; (SGR~1627)
Wachter et al.\ 2004; Corbel et al.\ 1999; (SGR~1801) Cline et al.\ 2000;
(SGR~1806) Kaplan et al.\ 2001; Fuchs et al.\ 1999; Corbel \& Eikenberry 2004;
(SGR~1900) Frail et al.\ 1999; Vrba et al.\ 2000; (CXO~0100) Lamb et al.\ 2002;
(4U~0142) Patel et al.\ 2003; Hulleman, van Kerkwijk \& Kulkarni 2004;
(1E~1048) Wang \& Chakrabarty 2002; (RXS~170849) Israel et al.\ 2003b;
(XTE~1810) Israel et al.\ 2004; (1E~1841) Wachter et al.\ 2004; Vasisht \&
Gotthelf 1997; (AX~1844) Vasisht et al.\ 2000; Gaensler et al.\ 1999; (1E~2259)
Patel et al.\ 2001; Kothes, Uyaniker \& Aylin 2002

}
\end{flushleft}

\end{center}

\end{minipage}
\end{table}

\subsection{SNR Associations}

Supernova remnants are the glowing relics of massive explosions produced during
the formation of some neutron stars.  The surface brightness of a SNR depends
upon its age and the density of the local inter-stellar medium.  Because of 
dimming and observational selection (see e.g.\ Gaensler \&
Johnston 1995), not every SNR contains a young pulsar.  Likewise, young pulsars
are not always found in SNRs.  In fact, only one third of young ($<$10$^{5}$
yr) pulsars are expected to be found positionally coincident with their
associated SNRs, which is entirely consistent with the observed fraction (\S7).

As shown by Gaensler et al.\ (2001), the situation is similar for the AXPs
and SGRs as a group.  Of the 10 confirmed magnetar candidates, only 2 have
solid SNR associations (\axpa\ with CTB~109 [Gregory \& Fahlman 1980] and
\axpe\ with Kes 73 [Vasisht \& Gotthelf 1997]).  The AXP candidate \axpg\ has
a solid association with the SNR G29.6$+$0.1 (Gaensler et al.\ 1999).  
Assuming that the SGR/AXP kick velocities incurred at birth are similar to
those of radio pulsars, and given the space density of SNRs in the LMC,
the association of \sgrb\ with the SNR N49 (Cline et al.\ 1982) was shown to be
less secure than previously thought:  the probability of a chance
alignment is about 0.5\% (Gaensler et al.\ 2001).  These authors argued
that other SGR/SNR and AXP/SNR associations reported in the literature were 
unconvincing, or likely to be chance superpositions.  

The ages of the remnants associated with the magnetar candidates are
$\sim$10$^{4}$ yr, consistent with other age estimates for most of these 
objects.  Other than containing a magnetar candidate, there is nothing
unusual about these two SNRs relative to those of comparable age which
are associated with radio pulsars.  Note also that the young 
characteristic age of SGR 1900$+$14 ($\sim 10^3$ yrs) would make the
absence of a SNR counterpart surprising, unless its true age were
significantly larger (Thompson et al. 2000).

\subsection{Galactic Distribution}

The distances to the SGRs and AXPs have been estimated in several several
different ways, with widely varying degrees of precision
(Table~\ref{tab:locations}). One SGR (\sgrb) and one AXP candidate (\axph) are
located in the Large and Small Magellanic Clouds, respectively, and have
well-determined distances.  Two AXPs (\axpa\ and \axpe) are positioned close to
the centers of SNR and are, very likely, physically associated (see
\S14.6.2).   Both \sgra\ and \sgrc\ may be  associated with massive star
clusters (see \S14.6.4), each of which has an estimated distance.  Most other
distances rely on the measurement of interstellar absorption from X-ray
spectra, and the intervening distribution of molecular clouds.  A more complete
discussion of the distance uncertainties is given in \"Ozel et al.\ (2001).

All except two of the magnetar candidates are located within our Galaxy, and
are positioned close to the Galactic plane.  Their estimated heights above (or
below) the Galactic plane for these sources are given in 
Table~\ref{tab:locations}.  The small rms scale height ($z_{\rm rms} \simeq$ 70
pc) implies  a young source population.  This is consistent with the young
ages  inferred from the spin parameters (Table~\ref{tab:timing}) and the SNR
associations.

\subsection{Radio Limits}

Despite deep, sensitive radio observations of most magnetar candidates, no
persistent radio emission has been detected from any SGR or AXP.  The only
recorded radio detection of a magnetar candidate was a transient outburst seen
from \sgrc\ in the days following the 1998 August 27 flare (see \S14.5.2). 
Pulsed emission from this SGR in 1998 near the X-ray pulse period (5.16 s) was
reported using 100 MHz data taken with the BSA (Shitov 1999), 
but the radio ephemeris disagreed significantly with the X-ray ephemeris 
(Woods et al.\ 1999c).  

The non-detections of magnetar
candidates at radio frequencies does not necessarily mean that they are very
different from standard pulsars in their radio properties
(Gaensler et al.\ 2001).
Most pulsars are expected to have fluxes below the current limits for the
AXPs.  Long-period pulsars tend to have narrower beams
($<$1 deg), effectively reducing the chances that their beams will cross our
line-of-sight.  Given the small number of observed SGRs and AXPs, it is not
surprising that none are detected.

The Parkes Multi-Beam Survey has revealed a handful of radio pulsars with
magnetar strength fields as inferred from their spin parameters (Camilo et al.\
2000; McLaughlin et al.\ 2003).  These pulsars have X-ray luminosities much
lower than their AXP/SGR counterparts (Pivovaroff, Kaspi \& Camilo 2000), in
spite of their similar dipole field strengths.  This result has led Pivovaroff
et al.\ and others (Camilo et al.\ 2000; McLaughlin et al.\ 2003) to conclude
that membership as an AXP (or SGR) requires more than just a strong dipole
magnetic field.  Nonetheless, it would not be surprising to observe
the transition of a high-field radio pulsar to a brighter X-ray state,
given the realization that some magnetar candidates can enter
X-ray low states for extended periods of time.

\subsection{Optical and IR Counterparts}

\begin{figure}[!htb]
\centerline{
\psfig{file=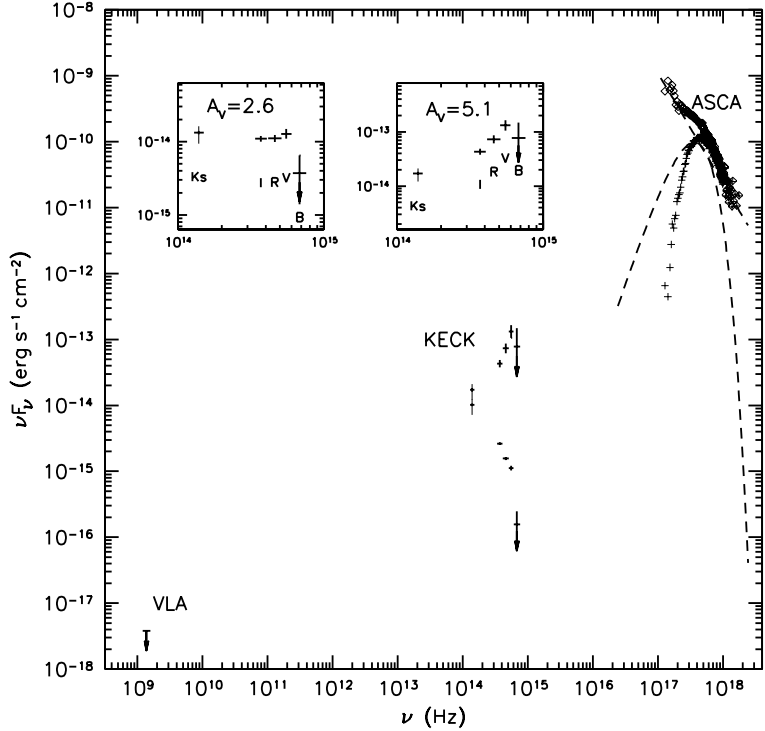,height=3.8in}}

\caption{Energy distribution for 4U~0142+61 (Hulleman et al.\ 2004).  At low
frequencies ($10^{14}$--$10^{15}$\,Hz), the points marked \texttt{V},
\texttt{R}, \texttt{I}, \texttt{K$_s$} indicate the observed 
V, R, I, and K$_s$-band fluxes.  The
vertical error bars reflect the uncertainties, while the horizontal ones
indicate the filter bandwidths.  The set of points above the measurements
indicate de-reddened fluxes for $A_V=5.4$, as inferred from the X-ray column
density. At high frequencies ($10^{17}$--$10^{18}$\,Hz), the crosses show the
incident X-ray spectrum as inferred from \texttt{ASCA} measurements.  The
diamonds show the spectrum after correction for interstellar absorption, and
the two thick dashed curves show the two components used in the fit.  Figure
courtesy of M. van Kerkwijk and F. Hulleman. \label{fig:nufnu_0142}}

\end{figure}

The first optical detection of a magnetar candidate, the AXP \axpc, 
was made by Hulleman et al.\ (2000).  They discovered an object with 
unusual colors spatially coincident with the X-ray position of the AXP.
The broadband spectrum of \axpc\ is shown in Figure~\ref{fig:nufnu_0142}.
Since this initial discovery, at least four other optical/IR counterparts
with similar characteristics have been discovered for other AXPs 
(and possibly one SGR).  The optical/IR properties of the magnetar 
candidates are given in Table~\ref{tab:opt_ir}.  See also Hulleman, 
van Kerkwijk \& Kulkarni (2004) for a table of empirically estimated
magnitudes of those sources not yet discovered.



\begin{table}[!h]
\begin{minipage}{1.0\textwidth}
\begin{center}
\caption{Optical and IR magnitudes of SGRs and AXPs. \label{tab:opt_ir}}
\label{tbl:optical}
\vspace{10pt}

\begin{tabular}{cccccccc}
\hline
Source                      &
V                           &
R                           &
I                           &  
J                           &  
H                           &  
K                           &  
K$_s$                       \\  

\hline

SGR~0526$-$66    & $>$27.1 &     ... &   $>$25 &     ... &     ... &     ... &     ...  \\
SGR~1627$-$41    &     ... &     ... &     ... & $>$21.5 & $>$19.5 &     ... & $>$20.0  \\
SGR~1806$-$20$^b$&     ... &     ... &     ... &   $>$21 & $>$20.5 &    18.6 &     ...  \\
SGR~1900$+$14    &     ... &     ... &     ... & $>$22.8 &     ... &     ... & $>$20.8  \\

\hline

CXOU~010043.1$-$721134 &     ... &    ...  &     ... &     ... &     ... &     ... &     ...  \\
4U~0142$+$61           &    25.6 &    25.0 &    23.8 &     ... &     ... &    19.6 &    20.1  \\
1E~1048.1$-$5937$^a$   &     ... & $>$24.8 &    26.2 &    21.7 &    20.8 &     ... & 19.4$-$21.3  \\
1RXS~J170849$-$400910$^b$&   ... &    ...  &     ... &    20.9 &    18.6 &     ... &    18.3  \\
XTE~J1810$-$197        &     ... &    ...  & $>$24.3 &     ... &    22.0 &     ... &    20.8  \\
1E~1841$-$045$^b$      &     ... & $>$23   &     ... &     ... &     ... &     ... &    19.4  \\
AX~J1844$-$0258        &     ... &    ...  &     ... &     ... &     ... &     ... &     ...  \\
1E~2259$+$586$^a$      &     ... & $>$26.4 & $>$25.6 & $>$23.8 &     ... &     ... & 20.4$-$21.7  \\

\hline

\end{tabular}

\begin{flushleft}
{\small 

$a$ -- Has shown variability in the infrared \\
$b$ -- Candidate counterpart \\

\vspace{11pt}

REFERENCES -- (SGR~0526) Kaplan et al.\ 2001; (SGR~1627) Wachter et al.\ 2004;
(SGR~1806) Eikenberry et al.\ 2001; (SGR~1900) Kaplan et al.\ 2002; (CXO~0100)
Lamb et al.\ 2002; (4U~0142) Hulleman et al.\ 2000, 2004; (1E~1048) Wang \&
Chakrabarty 2002; Israel et al.\ 2002; Durant, van Kerkwijk \& Hulleman 2003;
(RXS~170849) Israel et al.\ 2003b; (XTE~1810) Israel et al.\ 2004; (1E~1841)
Wachter et al.\ 2004, Mereghetti et al.\ 2001; (1E~2259) Hulleman et al.\ 2001;
Kaspi et al.\ 2003; Israel et al.\ 2003a

}
\end{flushleft}

\end{center}

\end{minipage}
\end{table}

The discovery of optical/IR counterparts to magnetar candidates has placed
valuable new constraints on their nature. 
Fast photometry of \axpc\ revealed optical ($R$ band) pulsations
with a high pulsed fraction ($\sim$27\% peak-to-peak) at the spin 
frequency of the neutron star (Kern \& Martin 2002).  The pulsed fraction
of the optical pulsations is comparable to or greater than
the pulsed fraction at X-ray energies.  

Monitoring of the optical/IR counterparts of the AXPs has shown that the IR
fluxes of the AXPs vary with time and burst activity (\S14.5.2). If the IR flux
is an indicator of burst emission in AXPs (e.g.\ \axpa\ in 2002 June), then IR
flux variability seen in \axpb\ (Israel et al.\ 2002) and \axpc\ (Hulleman, van
Kerkwijk \& Kulkarni 2004) could indicate the presence of undetected bursts. 
Clearly, continued monitoring of magnetar candidates is required to confirm
some of these early findings.

Deep IR observations of the fields of two SGRs have revealed the presence of
massive star clusters in which the SGRs are possibly embedded.  Fuchs et al.\
(1999) found that \sgra\ was positionally coincident with a cluster of
massive stars at a distance $\sim$15 kpc.  Indeed, a very luminous star is
located close to \sgra\ (van Kerkwijk et al. 1995), but is not positionally
coincident with it (Hurley et al.\ 1999d).  Similarly, Vrba et al.\ (2000) found
a cluster of massive stars at a distance $\sim$15 kpc surrounding \sgrc. 
Follow-up IR observations of \sgra\ have revealed two IR sources consistent
with the X-ray position of the SGR (Eikenberry et al.\ 2001), although the
ratio of optical to X-ray flux is anomalously large for either optical source,
thus neither are likely the counterparts to \sgra.  Analysis of the other
members of this cluster have shown that this particular cluster contains some
of the most massive stars in our Galaxy, perhaps even the most luminous star in
our Galaxy (Eikenberry et al.\ 2001).  These findings suggest that these two
SGRs may have very massive progenitors.

Finally, one other avenue open to pursue in optical/IR studies of magnetar
candidates is the measurement of proper motion.  Hulleman et al.\ (2000) have
already placed a 2$\sigma$ upper limit of 0.03 arc sec yr$^{-1}$ 
-- corresponding to $1400\,(D/10~{\rm kpc})$ km s$^{-1}$ --
on the proper motion of \axpc.  Some models of magnetar formation 
(e.g.\ Duncan \& Thompson 1992) 
suggest high kick velocities incurred at birth.  The precise astrometry
available with optical/IR observations will allow tighter constraints
on the proper motions of the AXPs and SGRs, and shed further light
on the physical connection of these sources to nearby SNR.

\section{Magnetar Model}

We will organize our discussion of the magnetar
model around its predictions, the extent to which they have been
verified or falsified, and outline areas in which further advancement of
theory is needed to make quantitative comparisons with data. 
The basic idea of the model is that the
variable X-ray emission -- the bursts lasting up to $\sim 1000$ s 
and the transient changes in persistent emission observed up to $\sim 1$ yr --
are powered by the decay of the star's magnetic field.  
A rms field exceeding $\sim 10^{15}$ G is needed to supply
an output of $10^{35}$ erg s$^{-1}$ extending over $10^4$ yrs.

We begin by recalling that the
AXPs as a group are systematically much brighter 
thermal X-ray sources than radio pulsars of the same characteristic age.
The evidence for
repeated bulk heating of the crust, and the relative brightness of
the X-ray emission, has bolstered the suggestion that magnetic field
decay is the main energy source for that emission.   That conclusion is
most secure in those magnetar candidates which show large transient swings
in X-ray brightness over a period of years.  Some SGRs and AXPs have
nearly flat 2-10 keV energy spectra, which manifestly cannot
be powered primarily by cooling or by
spin down.  Indeed, it has recently been found that the 
bolometric output of some AXPs is dominated by a hard, rising energy spectrum 
up to an energy of (at least) $\sim 100$ keV (Kuiper et al. 2004).

\subsection{Magnetic Field Decay}

Several physical effects become important as the magnetic field
of a neutron star is raised above $\sim 10^{14}$-$10^{15}$ G
(Thompson \& Duncan 1996).
First, a field stronger than $\sim 10^{15}$ G will, as it decays,
significantly raise the temperature of the 
deep crust and core of the star at an age of $\sim 10^3$-$10^4$ yrs.
The rate of drift of the magnetic field
and the entrained charged particles can, as a result, be significantly
accelerated.
Second, elastic stresses in the crust of the star are 
no longer able to withstand a large departure from magnetostatic
equilibrium.  The crustal lattice has a finite shear modulus $\mu$, and 
when the yield strain $\theta_{\rm max}$ is exceeded the lattice will
respond in an irreversible manner.  The characteristic magnetic field
strength is $B_{\rm yield} = (4\pi\theta_{\rm max} \mu)^{1/2}
= 2\times 10^{14}\,(\theta_{\rm max}/10^{-3})^{1/2}$ G.
For example, Hall drift of the magnetic field has qualitatively
different consequences in the crust when the field is stronger than
$B_{\rm yield}$.  The mean electron drift motion that supplies
the current also advects the magnetic field, and causes stresses
in the crust to slowly build up.  
When $B> B_{\rm yield}$, irregularities in the magnetic field
are damped directly by crustal yielding, rather than by
a non-linear coupling to high-frequency modes which suffer ohmic
damping.  Such a `Hall cascade' plays a key role in facilitating
the decay of $\sim 10^{12}$ G magnetic fields in the crust 
(Goldreich \& Reisenegger 1992).

A third, related effect is that when the star contains a strong
$\sim 10^{15}$ G toroidal magnetic field, the rate of ejection of 
magnetic helicity from the interior can be high enough to induce
a significant twist on the external poloidal field lines.  This
effect is especially important following periods of X-ray burst activity,
and has been suggested as a source of persistent increases in spin-down rate, 
and as a source of external heating (Thompson, Lyutikov, \& Kulkarni 2002).

The heat flux through the 
stellar crust is also modestly enhanced in strong magnetic fields
(van Riper 1988; Heyl \& Hernquist 1998; Potekhin \& Yakovlev 2001),
although the composition plays a more important role in determining
the thermal transparency.  In particular, the heat flux can be
up to several times larger if the surface has a light-element
(H or He) composition, than if it is iron
(Chabrier, Potekhin, \& Yakovlev 1997; Heyl \& Hernquist 1997).

Key early papers on the microscopic transport of
the magnetic field in neutron star interiors are by
Haensel, Urpin, \& Iakovlev (1990),
Goldreich \& Reisenegger (1992), and Pethick (1992).   Sweeping
of magnetic fluxoids out of a superconducting core by the
interaction with the superfluid vortices 
has been considered as a mechanism of magnetic field decay in
radio pulsars (Ruderman, Zhu, \&  Chen 1998).  It may, however,
be suppressed if the field is stronger than $\sim 10^{15}$ G
and the fluxoids are tightly bunched.  Integrated models of
magnetar evolution and cooling have been calculated by
Thompson \& Duncan (1996), Heyl \& Kulkarni (1998),
Colpi, Geppert, \& Page (2000), Kouveliotou et al. (2003).
Heating can have subtle
effects on the superfluid properties of the star: in particular,
if the critical temperature for the onset of neutron pairing
in the core is less than several $\times~10^8$~K, then it will
force a significant delay in the transition to core neutron superfluidity
(Arras, Cumming, \& Thompson 2004).  
After this transition, a neutron star undergoes a 
significant drop in surface X-ray flux (Yakovlev et al. 2001).

It should be emphasized that the
behavior of the SGRs and AXPs has not been observed over baselines 
longer than $\sim 20-30$ yrs.  Some AXPs such as \axpa\ and \axpc\ have 
sustained bright ($\sim 10^{35}$ erg s$^{-1}$) thermal X-ray emission
over this period of time, which is comparable to or longer than the
thermal conduction time across the crust (Gnedin et al. 2001).  Their
duty cycle as bright X-ray sources is presently unknown.

\subsection{Mechanism for Magnetar Bursts}

Both the short and the long outbursts of magnetars are hypothesized to arise
from the direct injection of energy into the magnetosphere, through a
rearrangement of the magnetic field and the formation and dissipation of strong
localized currents.  Magnetar flares are distinguished from Solar flares in two
key respects (Thompson \& Duncan 1995):   the magnetic field is anchored in a
rigid medium (the lower crust) which has some finite shear strength;  and the
energy density in the magnetic  field is high  enough that rapid thermalization
of energized charged particles can be expected.  (In the short SGR bursts, the
rate of release of energy is not rapid enough to effect complete
thermalization and drive the photon chemical
potential to zero, but it is at the onset of the giant flares;  
Thompson \& Duncan 2001.)

The crust provides a plausible site for the initial loss of equilibrium that
triggers an outburst.  For example,
the relaxation behavior observed over a period of weeks in SGR 1806-20 (Palmer
1999) suggests that the release of energy in successive short SGR bursts
is limited by inertial and frictional forces. 
In addition, bursts similar to short SGR bursts are
observed to begin at least two larger events:  the August 27 and August 29
flares of SGR 1900$+$14.  The duration of the short bursts is comparable to the
time for a torsional deformation  to propagate vertically across the crust
in a $\sim 10^{14}$ G poloidal magnetic field. The ability of the interior of
the star to store much stronger toroidal magnetic fields than the exterior
provides a hint that the ensuing burst is driven
primarily by a loss of equilibrium in the  crust, rather than by
reconnection and simplification of non-potential magnetic fields outside the
star.  Nonetheless, it is likely that both effects will occur in concert,
given the magnitude of the energies released.  

The initial spikes of the giant flares have been associated with
expanding fireballs composed of $e^\pm$ pairs and non-thermal gamma-rays
(Paczy\'nski 1992), and the pulsating tails
with thermalized energy which remains confined close to the neutron
star by its magnetic field (Thompson \& Duncan 1995).  In the spikes,
the combination of rapid ($< 0.01$ s) variability  with a hard
non-thermal spectrum points to a low baryon contamination.
The argument that most of the flare
energy is deposited in the first second comes from i) the near coincidence
between the energy of the initial spike and the energy radiated over
the remaining $\sim 300$ s of the burst; and ii)
the smooth adiabatic simplification of the pulse profile in the
tail of the 27 August 1998 flare, which shows no evidence for secondary
impulsive injections of energy that would be associated with a 
continuing substantial reorientation of the magnetic field.
The lower bound on the magnetic moment implied by the confinement of
$\sim 10^{44}$ ergs is $B R_{NS}^3 \simeq 10^{14}$G 
(Thompson \& Duncan 2001).  

Large-scale deformations of the crust are constrained by its
high hydrostatic pressure, but varying implications have been drawn for
its elastic response to evolving magnetic stresses.
One possibility is that the crust develops a dense network
of small-scale (but macroscopic) dislocations, and that the resulting 
fast ohmic heating of the uppermost layers of the star is what powers
the extended afterglow observed following SGR flares
(Lyubarsky, Eichler, \& Thompson 2002).  
Alternatively Jones (2003) and Lyutikov (2003) raise the possibility
that the response of the crust may be
more gradual and purely plastic, which would force the main source of 
energy for an X-ray flare into the magnetosphere.  
Evidence that the shear deformations
of the crust are spatially concentrated comes from
the observation of hard thermal X-ray emission -- covering $\sim 1$\% of 
the surface area of the star -- right after the August 29 flare 
of \sgrc\ (Ibrahim et al. 2001) and during the transient brightening 
of 1E 2259$+$586 (Woods et al. 2004).

\subsection{Burst Spectral Evolution and Afterglow}

A trapped thermal fireball (in which the photons have a Planckian 
distribution at a temperature $\sim 1$ MeV) is very optically
thick to scattering, given the high density of electron-positron pairs.
It releases energy through the contraction of its cool surface -- in
contrast to the cooling of a material body of fixed surface area.
Thus, the X-ray flux is predicted to drop rapidly toward the end of
a flare, when the external fireball evaporates (Thompson \& Duncan 1995).
A simple model of a contracting spherical surface, bounding
a fireball with a modest temperature gradient, provides an excellent 
fit to the 27 August 1998 flare (Feroci et al. 2001).

The temperature of the fireball surface is also buffered by a quantum
electrodynamic effect:  X-ray photons propagating through
intense magnetic fields are able to split in two or merge
together (Adler 1971). The rate of  splitting grows rapidly with 
photon frequency, but loses its dependence on magnetic field strength
when $B \gg B_{QED} = 4.4\times 10^{13}$ G (Thompson \& Duncan 1992).
Energy and momentum are both conserved in this process, with the 
consequence that only one polarization mode can split.  As a result,
splitting freezes out below a characteristic black body temperature of 
$\sim 12$ keV in super-QED magnetic fields (Thompson \& Duncan 1995).  
This is, very nearly, the temperature observed during an extended period 
of flux decline in the pulsating tail of the 
27 August 1998 flare (Feroci et al. 2001).  In some geometries, 
double Compton scattering can also be a significant source of photon
seeds near the scattering photosphere (Lyubarsky 2002).

The rate
of radiative conduction through an electron gas is greatly increased
by the presence of a strong magnetic field, which suppresses the opacity
of the extraordinary polarization mode (Sil'antev \&
Iakovlev 1980; Lyubarskii 1987).  Thus, the high luminosities of the 
intermediate flares, and the pulsating tails of the giant flares,
also point to the presence of $10^{14}-10^{15}$ 
G magnetic fields (Paczy\'nski 1992).
This effect can, however, be suppressed by mode exchange near the stellar
surface (Miller 1995), and probably requires a
confining magnetic field (Thompson \& Duncan 1995).  

One clear prediction of the trapped fireball model is that 
$\sim 1$ percent of the trapped energy will be conducted into the
surface of the neutron star over the duration of the fireball phase
(Thompson \& Duncan 1995).  
This energy can explain the prompt afterglow observed
immediately following the intermediate burst on 29 August 1998 
(Ibrahim et al. 2001), but heat conducted into the crust cannot
supply afterglow longer than $\sim 10^4$ s following the burst.
The relative importance of such conductive heating for the observed
afterglow -- as compared with direct bulk heating and continuing
relaxation of currents outside the star -- is not well understood.

A super-QED magnetic field has other interesting radiative effects.
The gyrational energy
of a proton or other ion can fall in the keV range:  $\hbar eB/m_pc$
$= 6.3\,(B/10^{15}~{\rm G})$ keV, possibly allowing for the formation
of absorption features (Zane et al. 2001; \"Ozel 2003; but see
Ho \& Lai 2003 for a discussion of how polarization mode switching
can drastically reduce the equivalent width of such a line feature).  
When the radiative flux out of
the star exceeds $\sim 10^{36}$ ergs s$^{-1}$, this means that
the radiative force applied at the cyclotron resonance can
exceed the force of gravity (Thompson et al. 2002).
The same large resonant cross-section also allows an ion component of
a persistent electric current flowing outside the star to have 
a measurable influence on the X-ray spectrum through cyclotron scattering.

Searches for X-ray lines during SGR bursts are potentially
diagnostic of the burst mechanism and the strength of the
magnetic field.  Short, low-energy SGR bursts are probably highly 
localized on the neutron star surface.  The magnetosphere is probably 
at a higher temperature than the surface during a burst, and so 
proton cyclotron features may be seen in emission in the keV range.

\subsection{Electrodynamics}

Highly non-thermal persistent X-ray emission is observed in the
actively bursting SGR sources, and from it one infers the presence of 
magnetospheric currents much stronger than the rotationally-driven 
Goldreich-Julian current.  Although the spin-down power in SGRs 
1806-20 and 1900+14 peaks at values approaching the X-ray
luminosity during their periods of most extreme spin-down torque, 
there is no correlation between the two.  To power
the X-ray emission, the energy that must be dissipated per 
Goldreich-Julian particle on the open field lines exceeds $\sim 10^8$ MeV.
By contrast, if the flux of particles close to the stellar surface
is normalized to $cB_{\rm NS}/4\pi eR_{\rm NS}$, then the energy
dissipated per particle need not exceed $\sim 100$ MeV (e.g., the binding
energy of an ion to the star).

The persistent changes in X-ray pulse profile observed following
SGR bursts could be caused by a change in the emission pattern; or
by a change in the distribution of particles which re-scatter the
X-rays higher in the magnetosphere (e.g.\ at $\sim 10$ neutron star
radii where the cyclotron resonance of the electrons is in the keV range). 
For example, in a magnetosphere threaded by persistent electric currents,
the optical depth at the cyclotron resonance of the current-carrying 
charges is of the order of unity over a continuous range of frequencies
(if the poloidal magnetic field is twisted through $\sim$1 radian; Thompson 
et al. 2002).  In such a situation, an X-ray photon will undergo a 
significant shift in frequency as it escapes the magnetosphere.

A non-thermal component of the X-ray spectrum is not always needed to fit the
persistent emission of the AXPs:  given the narrow bandpass being fit,
the convolution of two black bodies 
sometimes gives an acceptable fit in soft-spectrum sources  (Israel et al.
2000). Measurements of the AXP emission above several
keV are crucial to understanding the physical origin of the high energy excess.

\subsection{Torque Behavior}

In addition to the basic predictions of rapid spin down in the 
SGRs and bursting activity in the AXPs, theoretical work on magnetars
has anticipated some other observed properties of the magnetar
candidates.  The deduction that the magnetic fields of the SGRs are
time-variable suggested that the spin down would also be highly variable
(Thompson \& Blaes 1998).
If the magnetic field is variable on very short timescales, then
torque variations will arise from a continuous flux of high frequency
Alfv\'en waves and particles away from the star
(see also Harding, Contopoulos, and Kazanas 1999; Thompson et al. 2000).
This is the most plausible explanation for the
transient spin down observed in SGR 1900$+$14 at the time of the
27 August 1998 flare.  But the apparent time lag between bursting
activity and large torque variations in SGRs 1900$+$14 and 1806$-$20,
the observation of long-term ($>$ year)
and persistent increases in torque, and the persistence of the
changes in X-ray pulse profile following outbursts, are more 
consistent with the presence of large-scale currents on the closed
magnetospheric field lines (Thompson et al. 2002).
Torque variations could, in principle, also arise from a change
in the fraction of open field lines due to the suspension of a modest
amount of material inside the speed-of-light cylinder (Ibrahim et al. 2001);
but it is difficult to see why such material would not be redistributed
or expelled immediately following a bright SGR flare.

The magnetar model is conservative in the sense that the
AXPs and SGRs are assumed to be standard neutron stars, distinguished
only from radio pulsars (including the high-magnetic field tail of the
pulsar population) by the presence of a strong wound-up magnetic field
{\it inside} the star (Thompson \& Duncan 2001), and by the active
transfer of magnetic helicity across the stellar surface.
One way of testing this basic hypothesis is to 
search for glitches associated with a superfluid component.
Large glitches will be triggered in slowly rotating magnetars 
via the release of magnetic stresses in the crust -- either
due to sudden unpinning (Thompson \& Duncan 1993)
or to plastic deformations of the crust during which vortices remain
pinned (Thompson et al. 2000).  In the second case, spin down of
the superfluid occurs if the crust is twisted adiabatically about an axis 
that is tilted with respect to the rotation axis:  more superfluid
vortices move outward away from the rotation axis than move toward
it.  A large glitch observed in the AXP 1E 2259$+$586 (Kaspi et al. 2003;
Woods et al. 2004) provides a nice test of these ideas.  The year-long 
soft X-ray afterglow observed following the glitch suggests that the crust
was subject to a smooth, large-scale deformation.  Related effects
occurring in a superfluid core have also been implicated in the fast 
timing noise of the SGRs (Arras et al. 2004).

\section{Future Directions}

We close by outlining some major unsolved problems associated with
the SGRs and AXPs.

[1] {\it What is the birth rate of AXPs and SGRs compared with radio  pulsars? 
What fraction of neutron stars go through a phase of strong magnetic
activity?}  The selection of magnetar candidates -- through their burst
activity as SGRs, or through their persistent X-ray pulsations as AXPs -- is
limited by sensitivity.  There are $\sim$10 magnetar candidates in our Galaxy
and a conservative estimate of their average age is $\sim$10$^4$ years as
derived from their spin down.  Thus, a {\it lower limit} to the Galactic birth
rate is 1 per 1000 years (Kouveliotou et al.\ 1994; van Paradijs et al.\ 1995),
or $\sim$10\% of the radio pulsar birth rate (Lyne et al.\ 1998).  The birth
rate that we infer for AXPs and SGRs depends critically upon the efficiency
with which we detect them.   The efficiency of detecting low-luminosity bursts
such as those from \axpa\ is quite low.  The overall efficiency of identifying
magnetar candidates has not yet been quantified.  The observation of
transitions  to persistent low-luminosity states suggests that it may be lower
than previously thought; but these transient sources could be older on average
than the more persistent sources. Given the number of selection effects, we
cannot rule out a detection  efficiency as low as $\sim$10\%, and a birth rate
comparable to that of radio pulsars.

[2] {\it Why are the spin periods of AXPs and SGRs strongly clustered
in an interval of $5-12$ s?}  This clustering suggests a real upper
cutoff of $\sim$12 s in the period distribution (Psaltis \& Miller 2002).
In fact, the observed periods lie close to the upper envelope of the period 
distribution of radio pulsars, and are consistent with a reduction in torque
following the termination of active pair cascades on open magnetic
field lines (Thompson et al.\ 2002).  Field decay could, in principle,
also play a role in determining the observed range of spin periods
(Colpi et al.\ 2000).   The large $\sim 2\times 10^5$ yr
characteristic age of \axpa\ (which resides in the $\sim 10^4$-yr old
SNR CTB 109) provides a strong hint of torque decay in that particular AXP.

[3] {\it Do the SGRs, AXPs and high B-field radio pulsars form a continuum of
magnetic activity, or are they different phases/states of a more uniform 
class of object?} The heating of a neutron star by a decaying magnetic
field  is unfortunately sensitive to the configuration of the
field.   Arras et al. (2004) consider a toroidal configuration, 
and show that as the field strength is reduced from $\sim 10^{15}$ G 
down to  $\sim 10^{14}$ G, the soft X-ray luminosity  interpolates 
between the levels characteristic of
AXPs and of radio pulsars.  Thus the observed X-ray emission of middle-aged
radio pulsars is consistent with the hypothesis that the AXPs and SGRs have
much stronger internal magnetic fields.  However, a few AXPs have shown
transitions to low-luminosity states where their X-ray output is reduced by a
factor $\sim$100 on a time scale of years.  Much further exploration of the
interplay between magnetic field transport, surface cooling, and 
superfluidity is required.  For example, the relaxation of the crustal
magnetic field caused by electron captures on the heavy nuclei in the
neutron-drip solid has not yet been explored.

[4] {\it What can we learn of neutron star matter from observations of SGR
and AXP activity?  Are SGRs and AXPs fundamentally neutron stars?} 
The detection of glitches in two AXPs indicates the presence of
a superfluid component, whose pinning behavior (as deduced from
the post-glitch response) is similar to that observed in radio pulsars.  
Burst afterglows have the potential to probe the outer layers 
of magnetars. Lyubarsky
et al. (2002) argue that the extended afterglow observed following the
27 August 1998 flare is more consistent with the strongly stratified 
outer crust of a neutron star, than it is with a nearly constant density 
quark star.   It has been suggested that the high luminosities
of SGR flares are a result of QCD confinement near the surface of a bare
quark star (Usov 2001); but the theoretical motivation for such objects is
problematic (e.g., Akmal et al. 1998).

[5] {\it What is the initial spin period of magnetars? Could some magnetars
(with millisecond periods) be connected to GRBs?}  It is not known
whether magnetars and radio pulsars are distinguished by the initial
rotation of the neutron star, or alternatively by the stability properties 
of the magnetic field.  It is possible that all nascent neutron stars
develop $\sim 10^{15}$ G magnetic fields through fluid instabilities;
but it is also likely that the large-scale order of the magnetic
field is correlated with the speed of the rotation.  For example,
a large-scale helical dynamo is possible when the rotation
period is comparable to the timescale of the convective
motions, which is $\sim 3$ ms for Ledoux convection during the 
10-s Kelvin phase of the neutron star (Duncan \& Thompson 1992).  
This led to the prediction of a class of energetic supernovae in which
the neutron core deposits $\sim 10^{51}-10^{52}$ ergs of rotational energy 
by magnetic dipole radiation and later forms a strongly magnetic stellar
remnant.  The spin energy can be tapped even on the short timescale for the
shock to emerge from a compact CO core, when the effects of a neutrino-driven
wind are taken into account (Thompson, Chang \& Quataert 2004). Whether a 
proto-magnetar is also a viable source of gamma-ray burst emission 
(as suggested independently by
Usov 1992 and Duncan \& Thompson 1992) is more problematic:  the net
mass released during neutrino cooling is a few orders of magnitude 
larger than what will quench gamma-ray emission from the expanding
relativistic wind.

[6] {\it What is the evolutionary sequence of magnetars?  How do very young
($<$10$^3$ years) and older systems ($>$10$^5$ years) manifest themselves?
Are there old magnetars in our local neighborhood? in globular clusters?}
The limited $\sim 10^4$ yr lifetime of SGR flare activity is a significant
constraint on models of magnetic field decay.  This lifetime is determined
by the microscopic transport processes acting on the magnetic field; 
{\it and} by the manner in which the star falls out of magnetostatic
equilibrium  -- about which little is presently understood.  Hall drift
is not sensitive to temperature and can, for that reason, continue to power
a low level of X-ray emission ($\sim 10^{33}$ erg s$^{-1}$) at an age of 
$\sim 10^6$ yrs.  Rotational energy deposition also becomes more significant
at lower flux levels (e.g., Ruderman et al. 1998).
Evidence has recently been found for broad absorption features in the
spectra of a few soft-spectrum Dim Isolated Neutron Stars 
(van Kerkwijk et al.\ 2004; Haberl 2004)
perhaps indicative of proton cyclotron absorption in a 
magnetic field of several $\times 10^{13}$ G  (Zane et
al.\ 2001; Ho \& Lai 2003; \"Ozel 2003).  Their evolutionary
relation to radio pulsars and magnetars would be elucidated by a detailed
cross comparison with the low-luminosity states of magnetars.  Detailed
timing measurements are likely to be essential to unraveling these
interconnections.

[7] {\it Do magnetars exist in binary systems?  If so, how does a magnetar
react to accretion?  Or, alternatively, do magnetars form only through the
sacrifice of a binary companion?}   The modest number of binary neutron stars
which may have $10^{14}$ G magnetic fields (e.g., GX 1$+$4, 
2S 0114$+$650; Li \& van den Heuvel 1999) is probably
inconsistent with a magnetar birth exceeding $\sim 10$ percent of the total
neutron star birth rate, unless the magnetars have systematically larger
kicks or their dipole fields decay significantly above an age of 
$\sim 10^5$ yrs.  There is some evidence for large kicks in 
SGRs 0525$-$66 and 1900$+$14 (Cline et al.\ 1982; Thompson et al. 2000), but 
not in any of the AXPs.  The limited $\sim 10^4$ yr lifetime of AXPs
as bright X-ray sources makes it unlikely to see bright thermal
X-ray emission from magnetars formed in high-mass binaries.

\section{Acknowledgments}
We thank Shri Kulkarni for his help and advice in planning this review,
and for detailed comments on the manuscript.
We also thank Robert Duncan, Fotis Gavriil,
Marten van Kerkwijk, and Nanda Rea for their comments.


\begin{thereferences}

   \bibitem{Adler71}Adler, S.L., Ann. \ Phys. \ 67, 599

   \bibitem{aker00}Akerlof, C., et al. 2000, ApJ, 542, 251

   \bibitem{akmal98}Akmal, A., Pandharipande, V.~R., \& Ravenhall, D.~G.
    \ 1998, Phys. Rev. C, 58, 1804 

   \bibitem{AFO86}Alcock, C., Farhi, E. \& Olinto, A. 1986, Phys.\ Rev.\ 
   Lett.\ 57, 2088



   \bibitem{ABPC87}Ardelyan, N.~V., et al.
    1987, Soviet Astr., 31, 398

   \bibitem{konus01}Aptekar, R.L., et al.
   2001, ApJ, 137, 227

   \bibitem{Arras04}Arras, P., Cumming, A., \& Thompson, C. 2004, ApJ, 
   608, L49     

   \bibitem{arzou94}Arzoumanian, Z., Nice, D.J., Taylor, J.H. \& Thorsett,
   S.E. 1994, ApJ, 422, 671

   \bibitem{Atteia87}Atteia, J.L., et al. 1987, ApJ, 320, L105

   \bibitem{BZ34}Baade, W. \& Zwicky, F. 1934, Proc.\ Nat.\ Acad.\ Sci.\ 20,
   259

   \bibitem{BS96}Baykal, A. \& Swank, J. 1996, ApJ, 460, 470
   
   \bibitem{Bird04}Bird, A.J., et al. 2004, ApJ, 607, L33
   
   \bibitem{Blaes89}Blaes, O., Blandford, R.D., Goldreich, P., \& Madau
    P. 1989, ApJ, 343, 839

   \bibitem{Camilo00}Camilo, F., et al. 2000, ApJ, 541, 367

   \bibitem{Chabrier97} Chabrier, G., Potekhin, A.~Y., \& 
    Yakovlev, D.~G.\ 1997, ApJ, 477, L99 

   \bibitem{Chatterjee00a}Chatterjee, P., Hernquist, L., \& Narayan, 
   R.\ 2000, ApJ, 534, 373 

   \bibitem{Chatterjee00b}Chatterjee, P.~\& Hernquist, L.\ 2000, 
   ApJ, 543, 368 

   \bibitem{Cheng96}Cheng, B., Epstein, R.I., Guyer, R.A., \& Young, C. 1996,
   Nature, 382, 518

   \bibitem{Cline82}Cline, T.L., et al. 1982, ApJ, 255, L45

   \bibitem{Cline00}Cline, T.L., et al. 2000, ApJ, 531, 407

   \bibitem{Colpi00}Colpi, M., Geppert, U., \& Page, D.\ 2000, ApJ, 529, L29 


   \bibitem{Corbel99}Corbel, S., Chapuis, C., Dame, T.M., \& Durouchoux, P.
   1999, ApJ, 526, L29

   \bibitem{CE04}Corbel, S. \& Eikenberry, S.S. 2004, A\&A, 419, 191

   \bibitem{Corbet95}Corbet, R.H.D., et al.
   1995, ApJ, 443, 786

   \bibitem{dallosso03}Dall'Osso, S., et al.
   2003, ApJ, 499, 485

   \bibitem{hartog04}den Hartog, P., et al. 2004, ATEL 293
   
   \bibitem{DT92}Duncan, R. \& Thompson, C. 1992, ApJ, 392, L9
   
   \bibitem{Duncan01}Duncan, R. 2001, in 20$^{\rm th}$ Texas Symp. on
   Relativistic Astrophysics, Eds. J.C.~Wheeler \& H.~Martel, {\em AIP}, vol
   586, pp. 495

   \bibitem{durant03}Durant, M., van Kerkwijk, M.H., \& Hulleman, F. 2003, in
   Young Neutron Stars and their Environment: IAU Symp. 218, Eds. F.~Camillo    \& B.M.~Gaensler, p. 251 ({\tt astro-ph/0309801})


   \bibitem{Eik01}Eikenberry, S.S., et al. 2001, ApJ, 563, L133
   
   \bibitem{FG80}Fahlman G.G. \& Gregory, P.C. 1981, Nature, 293, 202

   \bibitem{Fenimore81}Fenimore, E.E., et al.
   1981, Nature, 289, 42

   \bibitem{Fenimore94}Fenimore, E., Laros, J.G., \& Ulmer, A. 1994, ApJ,
   432, 742

   \bibitem{Feroci99}Feroci, M., et al.
   1999, ApJ, 515, L9
   
   \bibitem{Feroci01}Feroci, M., Hurley, K., Duncan, R.C. \& Thompson, C.
   2001, ApJ, 549, 1021

   \bibitem{Feroci03}Feroci, M., et al.\ 2003, ApJ, 596, 470

   \bibitem{Feroci04}Feroci, M., et al.
   2004, ApJ, in press ({\tt/astro-ph/0405104})


   \bibitem{FKB99}Frail, D., Kulkarni, S. \& Bloom, J. 1999, Nature, 398, 127

   \bibitem{Fuchs99}Fuchs, Y., et al.
   1999, A\&A, 350, 891

   \bibitem{GS95}Gaensler, B.M. \& Johnston, S. 1995, MNRAS, 277, 1243

   \bibitem{GGV99}Gaensler, B.M., Gotthelf, E.V., \& Vasisht, G. 1999, ApJ,
   526, L37

   \bibitem{GSGV01}Gaensler, B.M., Slane, P.O., Gotthelf, E.V., \& Vasisht, G.
   2001, ApJ, 559, 963

   \bibitem{GK02}Gavriil, F. \& Kaspi, V.M. 2002, ApJ, 567, 1067

   \bibitem{GKW02}Gavriil, F.~P., Kaspi, V.~M., \& Woods, P.~M. 2002,
   Nature, 419, 142

   \bibitem{GKW04}Gavriil, F.~P., Kaspi, V.~M., \& Woods, P.~M. 2004,
   ApJ, 607, 959

   \bibitem{GK04}Gavriil, F. \& Kaspi, V.M. 2004, ApJ, 607, 959

   \bibitem{Giles96}Giles, A.B., et al. 1996, ApJ, 469, L25

   \bibitem{Gnedin01}Gnedin, O.~Y., Yakovlev, D.~G., \& Potekhin, A.~Y.\ 
    2001, MNRAS, 324, 725 

   \bibitem{Gogus99}{G\"o\u{g}\"u\c{s}}, E., et al.
   1999, ApJ, 526, L93

   \bibitem{Gogus01}{G\"o\u{g}\"u\c{s}}, E., et al.
   2001, ApJ, 558, 228

   \bibitem{Gogus02}{G\"o\u{g}\"u\c{s}}, E., et al.
   2002, ApJ, 577, 929

   \bibitem{Gold92}Goldreich, P. \& Reisenegger, A. 1992, ApJ, 395, 250

   \bibitem{Golen84}Golenetskii, S.V., Ilyinskii, V.N., Mazets, E.P. 1984,
   Nature, 307, 41


   \bibitem{Gott98}Gotthelf, E.V. \& Vasisht, G. 1998, New Astr., 3, 293

   \bibitem{Gott02}Gotthelf, E.V., et al.
   2002, ApJ, 564, L31

   \bibitem{Gott04}Gotthelf, E.V., Halpern, J.P., Buxton, M, \& Bailyn, C.
   2004, ApJ, 605, 368

   \bibitem{GF80}Gregory, P.C. \& Fahlman, G.G. 1980, Nature, 287, 805

   \bibitem{Guido04}Guidorzi, C., et al.
   2004, A\&A, 416, 297
   

   \bibitem{Haberl04}Haberl, F. 2004, Mem.~S.~A.~It., 75, 454
 
   \bibitem{Haensel90}Haensel, P., Urpin, V.A., \& Iakovlev, D.G. 1990,
   A\&A, 229, 133

   \bibitem{Harding99}Harding, A.K., Contopoulos, I., \& Kazanas, D. 1999, 
   ApJ, 525, L125

   \bibitem{Hellier94}Hellier, C. 1994, MNRAS, 271, L21

   \bibitem{Hewish68}Hewish, A., et al.
   1968, Nature, 217, 709

   \bibitem{HH97}Heyl, J.~S.~\& Hernquist, L.\ 1997, ApJ, 489, L67 

   \bibitem{HH98}Heyl, J.~S.~\& 
   Hernquist, L.\ 1998, MNRAS, 300, 599 

   \bibitem{HK98}Heyl, J.~S.~\&  Kulkarni, S.~R.\ 1998, ApJ, 506, L61 

   \bibitem{HH99}Heyl, J.S. \& Hernquist, L. 1999, MNRAS, 304, L37

   \bibitem{Ho03}Ho, W.C.G. \& Lai, D. 2003, MNRAS, 338, 233

   \bibitem{Hull01}Hulleman, F., et al. 2001, ApJ, 563, L49  

   \bibitem{HvKK00}Hulleman, F., van Kerkwijk, M.H. \& Kulkarni, S.R. 2000, 
   Nature, 408, 689  

   \bibitem{HvKK04}Hulleman, F., van Kerkwijk, M.H. \& Kulkarni, S.R. 2004, 
   A\&A, 416, 1037

   \bibitem{Hurley86}Hurley, K. 1986, Talk presented at the Gamma-Ray Stars
   Conference in Taos, NM

   \bibitem{Hurley94a}Hurley, K.J., McBreen, B., Rabbette, M., \& Steel, S. 
   1994, A\&A, 288, L49


   \bibitem{Hurley99a}Hurley, K., et al. 1999a, Nature, 397, 41   

   \bibitem{Hurley99b}Hurley, K., et al.
   1999b, ApJ, 510, L107
   
   \bibitem{Hurley99c}Hurley, K., et al. 1999c, ApJ, 510, L111

   \bibitem{Hurley99d}Hurley, K., et al. 1999d, ApJ, 523, L37

   \bibitem{Ibrahim01}Ibrahim, A., et al. 2001, ApJ, 558, 237

   \bibitem{ISP03}Ibrahim, A.I., Swank, J.H., \& Parke, W. 2003, ApJ, 584, L17

   \bibitem{Ibrahim04}Ibrahim, A.I., et al.\ 2004, ApJ, submitted,
   ({\tt astro-ph/0310665})

   \bibitem{Inan99}Inan, U.S., et al.
   1999, GeoRL, 26, 3357

   \bibitem{Israel00}Israel, G.L., et al. 2000, ApJ, 560, L65

   \bibitem{Israel02b}Israel, G.L., et al. 2002, ApJ, 580, L143

   \bibitem{Israel03a}Israel, G.L., et al. 2003a, in Young Neutron Stars and
   their Environment: IAU Symp. 218, Eds. F.~Camillo \& B.M.~Gaensler, in
   press, ({\tt astro-ph/0310482})

   \bibitem{Israel03b}Israel, G.L., et al. 2003b, ApJ, 589, L93

   \bibitem{Israel04}Israel, G.L., et al. 2004, ApJ, 603, L97

   \bibitem{IKH92}Iwasawa, K., Koyama, K. \& Halpern, J.P. 1992, Publ.
   Astron. Soc. Japan, 44, 9

   \bibitem{Jones03}Jones, P.~B.\ 2003, ApJ, 595, 342 

   \bibitem{Juett02}Juett, A.M., Marshall, H.L., Chakrabarty, D., \& Schulz,
   N.S. 2002, ApJ, 568, L31

   \bibitem{Kaplan01}Kaplan, D.L., et al. 2001, ApJ, 556, 399

   \bibitem{Kaplan02}Kaplan, D.L., Kulkarni, S.R., Frail, D.A., \& van
   Kerkwijk, M.H. 2002, ApJ, 566, 378

   \bibitem{Kaspi00}Kaspi, V.M., Lackey, J.R., \& Chakrabarty, D.  2000,
   ApJ, 537, L31

   \bibitem{Kaspi01}Kaspi, V.M., et al.
   2001, ApJ, 558, 253

   \bibitem{KGW03}Kaspi, V.M., et al.
   2003, ApJ, 588, L93

   \bibitem{KG03}Kaspi, V.M. \& Gavriil, F.P. 2003, ApJ, 596, L71

   \bibitem{KTU94}Katz, J.I., Toole, H.A., \& Unruh, S.H. 1994, ApJ, 437, 727

   \bibitem{KM02}Kern, B. \& Martin, C. 2002, Nature, 417, 527

   \bibitem{klose04}Klose, S., et al. 2004, ApJ in press 
   ({\tt astro-ph/0405299})

   \bibitem{koshut96}Koshut, T.M., et al. 1996, ApJ, 463, 570

   \bibitem{kothes02}Kothes, R., Uyaniker, B., \& Aylin, Y. 2002, ApJ, 576,
   169

   \bibitem{Kouveliotou87}Kouveliotou, C., et al. 1987, ApJ, 322, L21

   \bibitem{Kouveliotou94}Kouveliotou, C., et al. 1994, Nature, 368, 125

   \bibitem{Kouveliotou98a}Kouveliotou, C., et al. 1998a, Nature, 393, 235

   \bibitem{Kouveliotou98b}Kouveliotou, C., et al. 1998b, GCN Circ.\ 107

   \bibitem{Kouveliotou03}Kouveliotou, C., et al. 2003, ApJ, 596, L79
   
   \bibitem{KHN87}Koyama, K., Hoshi, R., \& Nagase, F. 1987, PASJ, 39, 801
   
   \bibitem{KHM04}Kuiper, L., Hermsen, W., \& Mendez, M. 2004, ApJ, 
   submitted, ({\tt astro-ph/0404582})

   \bibitem{Kulkarni03}Kulkarni, S.R., et al.
   2003, ApJ, 585, 948

   \bibitem{Lamb03a}Lamb, D., et al.\ 2003a, GCN Circ.\ 2351
    
   \bibitem{Lamb02}Lamb, R.C., Fox, D.W., Macomb, D.J., \& Prince, T.A. 2002, 
   ApJ, 574, L29

   \bibitem{Lamb03b}Lamb, D., Prince, T.A., Macomb, D.J., \& Majid, W.A. 2003b, 
   IAU Circ. 8220
    
   \bibitem{Laros87}Laros, J., et al. 1987, ApJ, 320, L111
    
   \bibitem{LW71}LeBlanc, J.M. \& Wilson, J.R. 1970, ApJ, 161, 541

   \bibitem{Lenters03}Lenters, G.T., et al. 2003, ApJ, 587, 761

   \bibitem{Lewin93}Lewin, W.H.G, van Paradijs, J., and Taam, R.E. 1993,
   Space Sci. Rev. 62, 223

   \bibitem{Li99}Li, X.-D. \& van den Heuvel, E.P.J. 1999, ApJ, 513, L45

   \bibitem{LT87}Livio, M. \& Taam, R.E. 1987, Nature, 327, 398

   \bibitem{Lyne98}Lyne, A.G., et al. 1998, MNRAS, 295, 743
    
   \bibitem{Lyuv97}Lyubarskii, Y.E. 1987, Astrophysics, 25, 277

   \bibitem{Lyubarsky02}Lyubarksy, Y. 2002, MNRAS, 332, 199

   \bibitem{LET03}Lyubarsky, E., Eichler, D., \& Thompson, C. 2002, ApJ, 580,
   L69

   \bibitem{Lyutikov03}Lyutikov, M.\ 2003, MNRAS, 346, 540 

   \bibitem{Manchester04}Manchester, R.N. 2004, Science, 304, 542

   \bibitem{MW01}Marsden, D. \& White, N.E. 2001, ApJ, 551, L155

   \bibitem{Mazets79}Mazets, E.P., et al.
   1979, Nature, 282, 587

   \bibitem{Mazets81}Mazets, E.P. \& Golenetskii, S.V. 1981, Ap\&SS, 75, 47

   \bibitem{Mazets99a}Mazets, E.P., et al.
   1999a, Astron. Lett., 25, 635

   \bibitem{Mazets99b}Mazets, E.P., et al.
   1999b,  ApJ, 519, L151

   \bibitem{Mclaugh03}McLaughlin, M.A., et al. 2003, ApJ, 591, L135

   \bibitem{MS95}Mereghetti, S. \& Stella, L. 1995, ApJ, 442, L17

   \bibitem{Mere00}Mereghetti, S., Cremonesi, D., Feroci, M., \& Tavani, M.
   2000, A\&A, 361, 240

   \bibitem{Mere01}Mereghetti, S., et al. 2001, MNRAS, 321, 143

   \bibitem{Mere04}Mereghetti, S., et al.
   2004, ApJ, in press, 
   ({\tt astro-ph/0404193})

   \bibitem{Miller95}Miller, M.C. 1995, ApJ, 448, L29

   \bibitem{Molkov04}Molkov, S.V., et al. 2004, Astr. Lett. in press, ({\tt
   astro-ph/0402416})
   
   \bibitem{Morii03}Morii, M., Sato, R., Kataoka, J., \& Kawai, N. 2003, PASJ,
   55, L45
   
   \bibitem{Mura94}Murakami, T., et al.
   1994, Nature, 368, 127

   \bibitem{Norris91}Norris, J.P., Hertz, P., Wood, K.S., \& Kouveliotou, C. 
   1991, ApJ, 366, 240

   \bibitem{olive02}Olive, J-F., et al. 2003, in Gamma-ray Burst and Afterglow
   Astronomy 2001, Eds. G.R.~Ricker \& R.K.~Vanderspek, {\em AIP}, vol 662, pp.
   82-87

   \bibitem{oost98}Oosterbroek, T., Parmar, A.N., Mereghetti, S., \& Israel, 
   G.L. 1998, A\&A, 334, 925

   \bibitem{Ozel02}\"Ozel, F. 2002, ApJ, 575, 397

   \bibitem{Ozel03}\"Ozel, F. 2003, ApJ, 583, 402

   \bibitem{Ozeletal01}\"Ozel, F., Psaltis, D. \& Kaspi, V.M. 2001, ApJ, 563,
   255

   \bibitem{Pacz90}Paczy\'nski, B., 1990, ApJ, 365, L9

   \bibitem{Pacz92}Paczy\'nski, B., 1992, Acta Astron., 42, 145

   \bibitem{Palmer99}Palmer, D.~M.\ 1999, ApJ, 512, L113 

   \bibitem{Palmer02}Palmer, D.M. 2002, in Soft Gamma Repeaters: The Rome
   2001 Mini-Workshop, Eds. M.~Feroci \& S.~Mereghetti, Mem.~S.~A.~It.,
   vol 73, n. 2, pp. 578-583

   \bibitem{Patel01}Patel, S.K., et al. 2001, ApJ, 563, L45

   \bibitem{Patel03}Patel, S.K., et al. 2003, ApJ, 587, 367

   \bibitem{perna01}Perna, R., Hernquist, L.E., \& Narayan, R. 2000, ApJ, 541,
   344

   \bibitem{Pethick92}Pethick, C.J. 1992, in Structure and Evolution 
   of Neutron Stars, eds. D. Pines, R. Tamagaki, and S. Tsuruta, p. 115

   \bibitem{PKC00}Pivovaroff, M.J., Kaspi, V.M., \& Camilo, F. 2000, ApJ, 535,
   379

   \bibitem{PY01}Potekhin, A.~Y.~\& Yakovlev, D.~G.\ 2001, A\&A, 374, 213 

   \bibitem{PM02}Psaltis, D. \& Miller, M.C. 2002, ApJ, 578, 325

   \bibitem{Ramaty80}Ramaty, R., et al.
   1980, Nature, 287, 122 

   \bibitem{Rea03}Rea, N., et al. 2003, ApJ, 586, L65

   \bibitem{Rev04}Revnivtsev, M.G., et al. 2004, Astr. Lett., 30(6), 382

   \bibitem{RKL94}Rothschild, R., Kulkarni, S. \& Lingenfelter, R. 1994,
   Nature, 368, 432

   \bibitem{Ruderman98}Ruderman, M., Zhu, T., \& Chen, K. 1998, ApJ, 492,
   267

   \bibitem{Shitov99}Shitov, Yu.P. 1999, IAU Circ. 7110

   \bibitem{Sym84}Symbalisty, E.M.D. 1984, ApJ, 285, 729

   \bibitem{SI80}Sil'antev, N.A. \& Iakovlev, D.G. 1980, Ap\&SS, 71, 45

   \bibitem{SI00}Strohmayer, T.E. \& Ibrahim, A.I. 2000, ApJ, 537, L111

   \bibitem{terrell80}Terrell, J., Evans, W.D., Klebesadel, R.W., \& Laros,
   J.G. 1980, Nature, 285, 383

   \bibitem{TB98}Thompson, C., \& Blaes, O. 1998, Phys. Rev. D, 57, 3219
   
   \bibitem{TD92}Thompson C. \& Duncan R.C. 1992, in
   Compton Gamma-Ray Observatory, eds. M. Friedlander,
   N. Gehrels and D.J. Macomb (AIP: New York), p. 1085
   
   \bibitem{TD93}Thompson, C., \& Duncan, R. 1993, ApJ, 408, 194

   \bibitem{TD95}Thompson, C., \& Duncan, R. 1995, MNRAS, 275, 255

   \bibitem{TD96}Thompson, C., \& Duncan, R. 1996, ApJ, 473, 322

   \bibitem{Thompson00}Thompson, C., et al.
   2000, ApJ, 543, 340

   \bibitem{TD01}Thompson, C., \& Duncan, R. 2001, ApJ, 561, 980

   \bibitem{TD02}Thompson, C., Lyutikov, M., \& Kulkarni, S.R.\ 2002, ApJ, 
   574, 332

   \bibitem{Thompson04}Thompson, T., Chang, P., \& Quataert, E. 2004,
   ApJ, in press ({\tt astro-ph/0401555})

   \bibitem{Torii98}Torii, K., Kinugasa, K., Katayama, K., \& Tsunemi, H. 1998,
   ApJ 503, 843

   \bibitem{Usov92}Usov, V.V. 1992, Nature, 357, 472

   \bibitem{Usov94}Usov, V.V. 1994, ApJ, 427, 984

   \bibitem{Usov01}Usov, V.~V.\ 2001, ApJ, 559, L135

   \bibitem{VK95}van Kerkwijk, M.~H., et al.
   1995, ApJ, 444, L33 

   \bibitem{VK04}van Kerkwijk, M.A., et al.
   2004, ApJ, in press ({\tt astro-ph/0402418})

   \bibitem{VTV95}van Paradijs, J., Taam, R.E. \& van den Heuvel, E.P.J.
   1995, A\&A, 299, L41

   \bibitem{vanRiper88}van Riper, K.~A.\ 1988, ApJ, 329, 339 

   \bibitem{Vasisht94}Vasisht, G., Kulkarni, S., Frail, D. \& Greiner, J. 
   1994, ApJ, 431, L35
      
   \bibitem{VG97}Vasisht, G. \& Gotthelf, E.V. 1997, ApJ, 486, L129

   \bibitem{Vasisht00}Vasisht, G., Gotthelf, E.V., Torii, K., \& Gaensler, B.M.
   2000, ApJ, 542, L49
      
   \bibitem{Vrba00}Vrba, F.J., et al.
   2000, ApJ, 533, L17

   \bibitem{Wachter04}Wachter, S., et al. 2004, ApJ, submitted, 
   ({\tt astro-ph/04mmnnn})

   \bibitem{WC02}Wang, Z. \& Chakrabarty, D. 2002, ApJ, 579, L33

   \bibitem{Woltjer64} Woltjer, L. 1964, ApJ, 140, 1309

   \bibitem{Woods99a}Woods, P.M., et al. 1999a, ApJ, 519, L139

   \bibitem{Woods99b}Woods, P.M., et al. 1999b, ApJ, 527, L47

   \bibitem{Woods99c}Woods, P.M., et al. 1999c, ApJ, 524, L55

   \bibitem{Woods00}Woods, P.M., et al. 2000, ApJ, 535, L55
   
   \bibitem{Woods01}Woods, P.M., et al. 2001, ApJ, 552, 748
   
   \bibitem{Woods02}Woods, P.M., et al. 2002, ApJ, 576, 381

   \bibitem{Woods03a}Woods, P.M., et al. 2003, ApJ, 596, 464
   

   \bibitem{Woods04a}Woods, P.M., et al. 2004, ApJ, 605, 378
   
   \bibitem{WW82}Woosley, S.E. \& Wallace, R.K. 1982, ApJ, 258, 716

   \bibitem{Yak01}Yakovlev, D.~G., Kaminker, A.~D., Gnedin, O.~Y., 
   \& Haensel, P.\ 2001, Phys. Rep., 354, 1 

   \bibitem{Zane01}Zane, S., Turolla, R., Stella, L, \& Treves, A. 2001,
   ApJ, 560, 384

\end{thereferences}
\end{document}